\documentclass[11pt,a4paper]{article}
\usepackage[dvips]{graphics}
\makeatletter \oddsidemargin -.1in \evensidemargin -.1in
\newcommand{\singlespacing}{\let\CS=\@currsize\renewcommand{\baselinestretch}{1}\tiny\CS}
\newcommand{\doublespacing}{\let\CS=\@currsize\renewcommand{\baselinestretch}{1.35}\tiny\CS}
\usepackage{graphics}
\usepackage{amsmath}
\usepackage{amssymb}
\usepackage{float, epsfig, floatflt, here}
\usepackage{cite}
\usepackage{longtable}
\usepackage{graphicx}
\usepackage{pdfpages}
\usepackage{makeidx}
\usepackage{caption}
\usepackage{subcaption}
\usepackage[export]{adjustbox}
\usepackage{quantikz}
\usepackage[section]{placeins}
\usepackage[pagewise]{lineno}
\usepackage{xcolor}
\usepackage{hyperref}
\title {\textbf{\LARGE  Mentor initiated Bi-directional Hybrid quantum Communication Protocol}}
\author{ {Manoj Kumar Mandal}$^{1}$ \thanks{e-mail: manojmandaliiest@gmail.com}, { Mitali Sisodia}$^{2}$ \thanks{e-mail: mitalisisodiyadc@gmail.com}, Plaban Saha $^{1}$ \thanks {e-mail : plabansaha.iiest@gmail.com},{ Binayak S. Choudhury}$^{1}$ \thanks {e-mail : binayak@math.iiests.ac.in}\\
~\\
$^{1}$ Department of Mathematics,
Indian Institute of Engineering Science and Technology,\\ Shibpur
B. Garden, Howrah - 711103,
 West Bengal, India\\
  $^{2}$ Indian Institute of Technology, Kanpur, Uttar Pradesh 208016  \\}
 
\begin{document}
\date{}\maketitle

\begin{abstract}
In this paper, we present a hybrid bidirectional controlled quantum communication protocol between two parties, initiated by a Mentor. Initially, the two main parties and the controller do not share a common quantum entanglement; instead, each party shares entanglement separately with the Mentor. The Mentor's actions create entanglement among the two parties and the controller. The protocol operates deterministically in the absence of environmental noise. Furthermore, we analyze the effects of various types of noises on the protocol, calculate the fidelity, and study how fidelity varies with different parameters. We also provide a theoretical description of the generation of the initial entangled channel used in the protocol. The quantum circuits for entanglement generation and also of the entire protocol are presented. To verify the protocol, these quantum circuits are executed on a real IBM quantum computer.
\end{abstract}

 {\textbf{Keywords:}} Mentor, Entanglement, Hybrid protocol, Quantum Teleportation, RSP, Quantum Noise.
\section{Introduction}
Quantum teleportation is one of the most fascinating and potentially transformative discoveries in modern physics. It is a groundbreaking concept in quantum mechanics that involves transferring the quantum state of a particle (such as a photon or an atom) from one location to another without physically moving the particle itself. It is based on the principles of quantum entanglement, a phenomenon where two particles become linked in such a way that the state of one particle directly affects the state of the other, regardless of the distance separating them. The discovery of quantum teleportation (QT) was done by Bennett et al. in 1993 for a single qubit quantum state \cite{teleportation}. In QT process and its variants (bidirectional QT, controlled QT, bidirectional CQT), the state to be teleported is unknown that can be seen in some recent references \cite{QTM, QT1, QT2, QT3, QT4, QT5, QT6, QT7} (and references therein). In the series of QT, another class of protocols i.e. remote state preparation (RSP) has been developed by H. K. Lo in 2000 \cite{RSP}, in which known quantum states are created at a distant location. RSP is a powerful technique for transmitting quantum states by utilizing pre-shared entanglement and classical communication, all while avoiding the need to physically transmit the states to the receiver, related work can be seen in papers \cite{RSP, RSP1, RSP2, RSP3, RSP4}. To perform communication between multiple parties with the RSP technique, two versions have been proposed i.e. joint remote state preparation (JRSP) and controlled remote state preparation (CRSP). JRSP refers to the process in quantum computing where multiple parties, each holding part of a quantum system, collaborate to prepare a shared quantum state  \cite{JRSP4, JRSP5, JRSP1, JRSP2, JRSP3}, while in CRSP a controller (third party) controls the RSP process, without controller's permission the state can not be prepared or task can not be completed \cite{CRSP1, CRSP2,CRSP3,CRSP4}. The cyclic process in CRSP can be found in references \cite{CRSP5, CRSP6,CRSP7,CRSP8}. 

There are certain Mentor initiated protocols in quantum communication studies which were introduced by Choudhury et al. in 2019 \cite{m1} and have been considered in works like \cite{m2, m3,m4}. A Mentor is an initiator of the protocol who, by certain acts of measurement, determines one of the many possible courses of the protocol. Also by his action the rest of the parties become connected through entanglement. The role of the Mentor is thereby finished and the Mentor quits the protocol. 

Later on, in advanced quantum communication, hybrid quantum information processing has been developed which refers to a network or an integrated system that combines different quantum communication technologies. A rapid development in the field of hybrid quantum communication, driven by both experimental advances and theoretical innovations can be seen in recent published works (2022-2024) - Mandal et al. \cite{H1, H6} proposed  bidirectional hybrid protocol which involves teleportation from one end and remote state preparation from the other end by using a five-qubit pure entangled state, Aharonov et al. \cite{H2} proposed a hybrid communication system that combines quantum cryptography with classical channels for error correction and synchronization, Smith et al. \cite{H3} presented a hybrid model that combines quantum repeaters with classical techniques for noise reduction and error correction and demonstrated how such hybrid systems can improve long-distance quantum communication by mitigating photon loss in optical fibers, Zor et al. \cite{H4} combined quantum key distribution (QKD) with quantum repeaters and classical communication protocols to demonstrate a scalable hybrid system over a 300 km fiber-optic link and showed the successful secure transmission, validating the feasibility of hybrid quantum networks over long distances.  In 2024, Cox \cite{H5} provided an extensive review of hybrid quantum communication protocols for the development of a quantum internet, Gong et al. presented controlled cyclic hybrid quantum communication \cite{H7, H8}. In 2023 \cite{H9}, Hua et al. proposed a hierarchical controlled hybrid quantum communication scheme utilizing a six-qubit entangled state,  Zhang et al. \cite{H10} worked on butterfly network coding which is based on hybrid controlled quantum communication.

The motivation to work on hybrid quantum communication arises from the published works mentioned above and from a combination of practical needs, technological advancements, and long-term goals for secure, efficient, and scalable communication networks. In this paper, we propose a hybrid bidirectional communication protocol supervised by a controller and initiated by a Mentor. This hybrid research involved the RSP and teleportation scheme which could not be completed without the controller. In a realistic scenario, without noise analysis, any research would not be practicable, the noise analysis part on each qubit is also discussed in this work. We have also verified the scheme and executed the circuit on a superconductivity-based IBM quantum computer.

The paper is organized as follows: In Section \ref {Main protocol}, the main hybrid protocol is discussed. Preparation of entangled channel is shown in Section \ref {Preparation of the entangled channels}. In Section \ref{Noise Analysis}, noise analysis on the proposed scheme is discussed. The protocol is executed on an IBM quantum computer and shown in Section \ref{IBM}. Finally, the results are concluded in the last section \ref{Conclusion}.

\section{Main protocol}\label{Main protocol}
Suppose there are two parties, Alice and Bob, located at two different distant locations. Alice wishes to teleport an unknown qubit in her possession to Bob, while Bob aims to create a known single-qubit state at Alice's location. Initially, Alice and Bob are not connected by an entangled channel. However, they are both linked to a third party, called Mentor, who initiates the protocol. Without Mentor's intervention, the protocol cannot be initiated or executed. Additionally, there is a Supervisor overseeing the protocol. The Supervisor's role is to provide the final signal required to complete the protocol, serving as a security mechanism. Without the Supervisor's action, the protocol cannot be concluded. All parties have access to classical communication channels, allowing them to exchange information as needed.\\

\noindent Suppose Alice wants to teleport a single qubit state $ |\psi_0\rangle_{A_0}=x|0\rangle+y|1\rangle$ to Bob and Bob wants to create a known state $ |\psi_1\rangle_{B_0}=a|0\rangle+b|1\rangle$ at the place of Alice where the coefficients satisfy the relations $|x|^2+|y|^2=1$ and $|a|^2+|b|^2=1$. Two entangled channels, one between Alice and Mentor, and another one between Mentor Bob and Controller, are given respectively as
\begin{equation}
\begin{split}
    |\Xi_1\rangle_{M-A}=&\frac{1}{2}(|0000\rangle+|0101\rangle+|1010\rangle-|1111\rangle)_{m_1m_2A_1A_2}\\
     |\Xi_2\rangle_{M-B-C}=&\frac{1}{2}(|00000\rangle+|01011\rangle+|10101\rangle-|11110\rangle)_{m_3 m_4 B_1 B_2 C}
\end{split}
\end{equation}
where the qubits $m_1, m_2, m_3$, and $m_4$ belong to the Mentor and qubits $A_1$ and $A_2$ are in the hands of Alice, Bob possess the qubits $B_1$ and $B_2$ and the qubit $C$ belongs to the Controller.
The steps of the protocol are described as\\

\noindent\textbf{Step 1:} Mentor makes two Bell basis measurement on the qubit pairs $(m_1, m_3)$, $(m_2, m_4)$ and shares his result with the other parties. The action of the Mentor creates entanglement between the other three parties, Alice, Bob, and the Controller. There will be 16 possible cases which will be fixed by Mentor's measurement. This is the end of the Mentor's role in the protocol. \\
Bell basis is given by the following four linearly independent entangled states 
\begin{equation}\label{eq2}
    \begin{split}
        |\Phi_0\rangle=\frac{1}{\sqrt{2}}(|00\rangle+|11\rangle),~~~ & |\Phi_1\rangle=\frac{1}{\sqrt{2}}(|00\rangle-|11\rangle)\\
         |\Phi_2\rangle=\frac{1}{\sqrt{2}}(|01\rangle+|10\rangle),~~~& |\Phi_3\rangle=\frac{1}{\sqrt{2}}(|01\rangle-|10\rangle)
    \end{split}
\end{equation}
\noindent\textbf{Step 2:} Alice makes a Bell basis measurement given in equation (\ref{eq2}) on the qubit pair $(A_0, A_1)$ and shares his result with Bob and Controller classically.\\

\noindent\textbf{Step 3:} Bob makes a measurement on his qubit $B_2$ with the basis $\{|\xi_0\rangle, |\xi_1\rangle\}$ and announces his outcome via classical channel. The basis $\{|\xi_0\rangle, |\xi_1\rangle\}$ is given by
\begin{equation}\label{eq3}
    \begin{split}
        |\xi_0\rangle=a|0\rangle+b|1\rangle,~~~ & |\xi_1\rangle=b|0\rangle-a|1\rangle
    \end{split}
\end{equation}
This basis construction is possible only by Bob since alone he has complete knowledge about $a$ and $b$, and no other parties have the information regarding $a$ and $b$.\\

\noindent\textbf{Step 4:}  After getting the information of the measurement outcomes by the other parties, the Controller analyzes the output and the performance of all the parties and if he feels satisfied, makes a measurement on his qubit $C$ in the X-basis $\{|+\rangle, |-\rangle\}$ and shares his result with Alice and Bob. 
\begin{equation}
    \begin{split}
        |+\rangle=\frac{1}{\sqrt{2}}(|0\rangle+|1\rangle),~~~ & |-\rangle=\frac{1}{\sqrt{2}}(|0\rangle-|1\rangle)
    \end{split}
\end{equation}
\noindent\textbf{Step 5:} Finally Alice and Bob apply appropriate unitary operations according to the measurement outcomes of the other parties on their respective qubits $A_2$ and $B_1$ to get the desired state. All the unitary operators are listed in Table \ref{tab:my_label_B} and Table \ref{tab:my_label_A}. This is the end of the protocol.\\

\noindent Now we will describe the protocol in detail.\\

\noindent The combined state of the two entangled channels is given by
\begin{equation}
\begin{split}
|\tau\rangle= &|\Xi_1\rangle_{M-A}\otimes  |\Xi_2\rangle_{M-B-C}\\
    =&\frac{1}{4}\big[|0000 00000\rangle+|0001 00011\rangle+|0100 00101\rangle-|0101 00110\rangle\\
    &+|0010 01000\rangle+|0011 01011\rangle+|0110 01101\rangle-|0111 01110\rangle\\
    &+|1000 10000\rangle+|1001 10011\rangle+|1100 10101\rangle-|1101 10110\rangle\\
    &-|1010 11000\rangle-|1011 11011\rangle-|1110 11101\rangle+|1111 11110\rangle\big]_{m_1m_3m_2 m_4A_1A_2B_1 B_2 C} \\
\end{split}
\end{equation}
 The above combined entangled states can be expressed using the Bell basis given in equation (2) as
\begin{equation}
|\tau\rangle=\frac{1}{4}\sum_{i=0}^{3}\sum_{j=0}^{3}|\Phi_i\rangle_{m_1m_3}\otimes|\Phi_j\rangle_{m_2m_4}\otimes |M_{ij}\rangle_{A_1A_2B_1 B_2 C}
\end{equation}
where the states $|M_{ij}\rangle_{A_1A_2B_1 B_2 C}$ are given by
\begin{center}
     $|M_{00}\rangle=\frac{1}{2}[|00000\rangle+|01011\rangle+|10101\rangle+|11110\rangle\big]_{A_1A_2B_1 B_2 C}$ \\
     $|M_{01}\rangle=\frac{1}{2}[|00000\rangle-|01011\rangle+|10101\rangle-|11110\rangle\big]_{A_1A_2B_1 B_2 C}$\\
     $|M_{02}\rangle=\frac{1}{2}[|00011\rangle+|01000\rangle-|10110\rangle-|11101\rangle\big]_{A_1A_2B_1 B_2 C}$\\
     $|M_{03}\rangle=\frac{1}{2}[|00011\rangle-|01000\rangle-|10110\rangle+|11101\rangle\big]_{A_1A_2B_1 B_2 C}$\\
     $|M_{10}\rangle=\frac{1}{2}[|00000\rangle+|01011\rangle-|10101\rangle-|11110\rangle\big]_{A_1A_2B_1 B_2 C}$\\
     $|M_{11}\rangle=\frac{1}{2}[|00000\rangle-|01011\rangle-|10101\rangle+|11110\rangle\big]_{A_1A_2B_1 B_2 C}$\\
     $|M_{12}\rangle=\frac{1}{2}[|00011\rangle+|01000\rangle+|10110\rangle+|11101\rangle\big]_{A_1A_2B_1 B_2 C}$\\
     $|M_{13}\rangle=\frac{1}{2}[|00011\rangle-|01000\rangle+|10110\rangle-|11101\rangle\big]_{A_1A_2B_1 B_2 C}$\\
     $|M_{20}\rangle=\frac{1}{2}[|00101\rangle-|01110\rangle+|10000\rangle-|11011\rangle\big]_{A_1A_2B_1 B_2 C}$ \\
     $|M_{21}\rangle=\frac{1}{2}[|00101\rangle+|01110\rangle+|10000\rangle+|11011\rangle\big]_{A_1A_2B_1 B_2 C}$\\
     $|M_{22}\rangle=\frac{1}{2}[-|00110\rangle+|01101\rangle+|10011\rangle-|11000\rangle\big]_{A_1A_2B_1 B_2 C}$\\
     $|M_{23}\rangle=\frac{1}{2}[-|00110\rangle-|01101\rangle+|10011\rangle+|11000\rangle\big]_{A_1A_2B_1 B_2 C}$\\
     $|M_{30}\rangle=\frac{1}{2}[|00101\rangle-|01110\rangle-|10000\rangle+|11011\rangle\big]_{A_1A_2B_1 B_2 C}$\\
     $|M_{31}\rangle=\frac{1}{2}[|00101\rangle+|01110\rangle-|10000\rangle-|11011\rangle\big]_{A_1A_2B_1 B_2 C}$ \\
     $|M_{32}\rangle=\frac{1}{2}[-|00110\rangle+|01101\rangle-|10011\rangle+|11000\rangle\big]_{A_1A_2B_1 B_2 C}$\\
     $|M_{33}\rangle=\frac{1}{2}[-|00110\rangle-|01101\rangle-|10011\rangle-|11000\rangle\big]_{A_1A_2B_1 B_2 C}$
\end{center}
The Mentor performs two Bell basis measurements on the qubit pairs $(m_1, m_3)$ and $(m_2, m_4)$ and communicates the outcomes of these measurements to Alice, Bob, and the Controller via classical channels. If the measurement outcomes are $|\Phi_i\rangle_{m_1m_3}$ and $|\Phi_j\rangle_{m_2m_4}$, the reduced state of the remaining qubits $A_1, A_2, B_1, B_2$, and $C$ becomes $|M_{ij}\rangle_{A_1A_2B_1B_2C}$ which is an entangled resource shared among Alice, Bob, and the Controller, forming the foundation for the subsequent steps of the protocol.\\

\noindent Suppose Mentor's measurement result are $|\Phi_0\rangle_{m_1m_3}$, and $|\Phi_1\rangle_{m_2m_4}$. Then the  reduced state of the qubits $A_1, A_2, B_1, B_2$ and $C$ is $|M_{01}\rangle_{A_1A_2B_1 B_2 C}$ which is an entangled state. Now the   whole system  with the known qubit $|\psi_0\rangle_{A_0}$ is given by
\begin{equation}
    \begin{split}
        |\Phi\rangle=&(x|0\rangle+y|1\rangle)_{A_0}\otimes  \frac{1}{2}(|00000\rangle-|01011\rangle+|10101\rangle-|11110\rangle\big)_{A_1A_2B_1 B_2 C}
    \end{split}
\end{equation}
We can be expressed the above-combined state $ |\Phi\rangle$  in the basis given in equations (2), (3) and (4) as
\begin{equation}
    \begin{split}
        |\Phi\rangle=&\frac{1}{4}|\Phi_0\rangle_{A_0 A_1}\otimes|\xi_0\rangle_{B_2} \otimes|+\rangle_C\otimes[xa|00\rangle-xb|10\rangle+ya|01\rangle-yb|11\rangle\big]_{A_2B_1}\\
                   +&\frac{1}{4}|\Phi_0\rangle_{A_0 A_1}\otimes|\xi_0\rangle_{B_2} \otimes|-\rangle_C\otimes[xa|00\rangle+xb|10\rangle-ya|01\rangle-yb|11\rangle\big]_{A_2B_1}\\
                   +&\frac{1}{4}|\Phi_0\rangle_{A_0 A_1}\otimes|\xi_1\rangle_{B_2} \otimes|+\rangle_C\otimes[xb|00\rangle+xa|10\rangle+yb|01\rangle+ya|11\rangle\big]_{A_2B_1}\\ 
                   +&\frac{1}{4}|\Phi_0\rangle_{A_0 A_1}\otimes|\xi_1\rangle_{B_2} \otimes|-\rangle_C\otimes[xb|00\rangle-xa|10\rangle-yb|01\rangle+ya|11\rangle\big]_{A_2B_1}\\               
                   +&\frac{1}{4}|\Phi_1\rangle_{A_0 A_1}\otimes|\xi_0\rangle_{B_2} \otimes|+\rangle_C\otimes[xa|00\rangle-xb|10\rangle-ya|01\rangle+yb|11\rangle\big]_{A_2B_1}\\
                   +&\frac{1}{4}|\Phi_1\rangle_{A_0 A_1}\otimes|\xi_0\rangle_{B_2} \otimes|-\rangle_C\otimes[xa|00\rangle+xb|10\rangle+ya|01\rangle+yb|11\rangle\big]_{A_2B_1}\\                  
                    +&\frac{1}{4}|\Phi_1\rangle_{A_0 A_1}\otimes|\xi_1\rangle_{B_2}\otimes|+\rangle_C\otimes [xb|00\rangle+xa|10\rangle-yb|01\rangle-ya|11\rangle\big]_{A_2B_1}\\
                    +&\frac{1}{4}|\Phi_1\rangle_{A_0 A_1}\otimes|\xi_1\rangle_{B_2}\otimes|-\rangle_C\otimes [xb|00\rangle-xa|10\rangle+yb|01\rangle-ya|11\rangle\big]_{A_2B_1}\\
                    +&\frac{1}{4}|\Phi_2\rangle_{A_0 A_1}\otimes|\xi_0\rangle_{B_2} \otimes|+\rangle_C\otimes[ya|00\rangle-yb|10\rangle+xa|01\rangle-xb|11\rangle\big]_{A_2B_1}\\
                     +&\frac{1}{4}|\Phi_2\rangle_{A_0 A_1}\otimes|\xi_0\rangle_{B_2} \otimes|-\rangle_C\otimes[ya|00\rangle+yb|10\rangle-xa|01\rangle-xb|11\rangle\big]_{A_2B_1}\\
                    +&\frac{1}{4}|\Phi_2\rangle_{A_0 A_1}\otimes|\xi_1\rangle_{B_2} \otimes|+\rangle_C\otimes[yb|00\rangle+ya|10\rangle+xb|01\rangle+xa|11\rangle\big]_{A_2B_1}\\
                    +&\frac{1}{4}|\Phi_2\rangle_{A_0 A_1}\otimes|\xi_1\rangle_{B_2} \otimes|-\rangle_C\otimes[yb|00\rangle-ya|10\rangle-xb|01\rangle+xa|11\rangle\big]_{A_2B_1}\\
                    +&\frac{1}{4}|\Phi_3\rangle_{A_0 A_1}\otimes|\xi_0\rangle_{B_2} \otimes|+\rangle_C\otimes[-ya|00\rangle+by|10\rangle+xa|01\rangle-xb|11\rangle\big]_{A_2B_1}\\
                    +&\frac{1}{4}|\Phi_3\rangle_{A_0 A_1}\otimes|\xi_0\rangle_{B_2} \otimes|-\rangle_C\otimes[-ya|00\rangle-by|10\rangle-xa|01\rangle-xb|11\rangle\big]_{A_2B_1}\\  
    \end{split}
\end{equation}
\begin{equation*}
    \begin{split}
                    +&\frac{1}{4}|\Phi_3\rangle_{A_0 A_1}\otimes|\xi_1\rangle_{B_2}\otimes|+\rangle_C\otimes [-yb|00\rangle-ya|10\rangle+xb|01\rangle+xa|11\rangle\big]_{A_2B_1}\\
                    +&\frac{1}{4}|\Phi_3\rangle_{A_0 A_1}\otimes|\xi_1\rangle_{B_2}\otimes|-\rangle_C\otimes [-yb|00\rangle+ya|10\rangle-xb|01\rangle+xa|11\rangle\big]_{A_2B_1}\\
    \end{split}
\end{equation*}
Now Alive and Bob make their measurements as described in the step 2 and step 3. Suppose Alice's measurement result is $|\Phi_0\rangle_{A_0A_1}$ and Bob's measurement outcome is $|\xi_1\rangle_{B_2}$ then the reduced state of the qubits $A_2, B_1$ and $C$ is given by
\begin{equation}
    \begin{split}
        |Y_1\rangle=&\frac{1}{\sqrt{2}} |+\rangle_C\otimes[xb|00\rangle+xa|10\rangle+yb|01\rangle+ya|11\rangle\big]_{A_2B_1}\\ 
                   +&\frac{1}{\sqrt{2}}|-\rangle_C\otimes[xb|00\rangle-xa|10\rangle-yb|01\rangle+ya|11\rangle\big]_{A_2B_1}
    \end{split}
\end{equation}
After being satisfied with the performance of all the parties, the Controller measures qubit $C$ on the basis described in step 4 and announces his outcome classically. Since qubit $C$ is entangled with the qubits $A_2$ and $B_1$, Alice and Bob would not have reached the desired state without his action. \\

\noindent If Controller's measurement result is $|+\rangle_C$ then the reduced state of the qubits $A_2$, and $B_1$ is
\begin{equation}
    \begin{split}
        |Y_2\rangle=[xb|00\rangle+xa|10\rangle+yb|01\rangle+ya|11\rangle\big]_{A_2B_1}=(b|0\rangle+a|1\rangle)_{A_2}\otimes(x|0\rangle+y|1\rangle)_{B_1}
    \end{split}
\end{equation}
In this case Alice needs to apply unitary operation $\sigma_x$ on her qubit $A_2$ to get the desired state that Bob wants to create and Bob needs to operate $I$ on his qubit $B_1$ to get the desired state that Alice wants to teleport.\\

\noindent If Controller's measurement result is $|-\rangle_C$ then the reduced state of the qubits $A_2$, and $B_1$ is
\begin{equation}
    \begin{split}
        |Y_3\rangle=[xb|00\rangle-xa|10\rangle-yb|01\rangle+ya|11\rangle\big]_{A_2B_1}=(b|0\rangle-a|1\rangle)_{A_2}\otimes(x|0\rangle-y|1\rangle)_{B_1}
    \end{split}
\end{equation}
In this scenario, Alice must apply the unitary operation $\sigma_x\sigma_z $ ( as in table 2) on her qubit $A_2$ to achieve the desired state that Bob intends to create. Meanwhile, Bob needs to apply the identity operation  $\sigma_x$ ( as in table 1) on his qubit $B_1$ to obtain the desired state that Alice wishes to teleport.\\

\noindent All other possible cases and their corresponding unitary operations are given in Table \ref{tab:my_label_B} and Table \ref{tab:my_label_A}. This is the end of the protocol.
\begin{table}[ht]
    \centering
    \begin{tabular}{|c|c|c|c|c|}
    \hline
    Mentor's&Alice's & Controller's & State of &Unitary operation \\
   Outcomes& Outcomes&  Outcomes&  Bib's qubit ($B_1$)& by Bob $(U_{Bob})$\\
    \hline
         &$|\Phi_0\rangle_{A_0A_1}$& $|+\rangle_C$ &$ x|0\rangle+y|1\rangle$& $I$ \\
         & $|\Phi_0\rangle_{A_0A_1}$& $|-\rangle_C$ &$ x|0\rangle-y|1\rangle$& $\sigma_z$ \\
           $\{(|\Phi_0\rangle, |\Phi_0\rangle),$&$|\Phi_1\rangle_{A_0A_1}$& $|+\rangle_C$ &$ x|0\rangle-y|1\rangle$& $\sigma_z$ \\
          $(|\Phi_0\rangle, |\Phi_1\rangle),$&$|\Phi_1\rangle_{A_0A_1}$& $|-\rangle_C$ &$ x|0\rangle+y|1\rangle$& $I$ \\
           $(|\Phi_1\rangle, |\Phi_2\rangle),$ & $|\Phi_2\rangle_{A_0A_1}$& $|+\rangle_C$ &$ y|0\rangle+x|1\rangle$& $\sigma_x$ \\
         $(|\Phi_1\rangle, |\Phi_3\rangle)\}$ &$|\Phi_2\rangle_{A_0A_1}$& $|-\rangle_C$ &$ y|0\rangle-x|1\rangle$& $\sigma_x\sigma_z$ \\
          &$|\Phi_3\rangle_{A_0A_1}$& $|+\rangle_C$ &$ -y|0\rangle+x|1\rangle$& $\sigma_z\sigma_x$ \\
          &$|\Phi_3\rangle_{A_0A_1}$& $|-\rangle_C$ &$ -y|0\rangle-x|1\rangle$& $\sigma_z\sigma_x\sigma_z$ \\
         \hline
           &$|\Phi_0\rangle_{A_0A_1}$& $|+\rangle_C$ &$ x|0\rangle-y|1\rangle$& $\sigma_z$ \\
         & $|\Phi_0\rangle_{A_0A_1}$& $|-\rangle_C$ &$ x|0\rangle+y|1\rangle$& $I$ \\
           $\{(|\Phi_1\rangle, |\Phi_0\rangle),$&$|\Phi_1\rangle_{A_0A_1}$& $|+\rangle_C$ &$ x|0\rangle+y|1\rangle$& $I$ \\
         $(|\Phi_1\rangle, |\Phi_1\rangle),$ &$|\Phi_1\rangle_{A_0A_1}$& $|-\rangle_C$ &$ x|0\rangle-y|1\rangle$& $\sigma_z$ \\
          $(|\Phi_0\rangle, |\Phi_2\rangle),$  & $|\Phi_2\rangle_{A_0A_1}$& $|+\rangle_C$ &$ y|0\rangle-x|1\rangle$& $\sigma_x\sigma_z$ \\
         $(|\Phi_0\rangle, |\Phi_3\rangle)\}$ &$|\Phi_2\rangle_{A_0A_1}$& $|-\rangle_C$ &$ y|0\rangle+x|1\rangle$& $\sigma_x$ \\
          &$|\Phi_3\rangle_{A_0A_1}$& $|+\rangle_C$ &$ -y|0\rangle-x|1\rangle$& $\sigma_z\sigma_x\sigma_z$ \\
          &$|\Phi_3\rangle_{A_0A_1}$& $|-\rangle_C$ &$ -y|0\rangle+x|1\rangle$& $\sigma_z\sigma_x$ \\
         \hline
         &$|\Phi_0\rangle_{A_0A_1}$& $|+\rangle_C$ &$ y|0\rangle+x|1\rangle$& $\sigma_x$ \\
         & $|\Phi_0\rangle_{A_0A_1}$& $|-\rangle_C$ &$ y|0\rangle-x|1\rangle$& $\sigma_x\sigma_z$ \\
           $\{(|\Phi_2\rangle, |\Phi_0\rangle),$&$|\Phi_1\rangle_{A_0A_1}$& $|+\rangle_C$ &$ -y|0\rangle+x|1\rangle$& $\sigma_z\sigma_x$ \\
          $(|\Phi_2\rangle, |\Phi_1\rangle),$&$|\Phi_1\rangle_{A_0A_1}$& $|-\rangle_C$ &$ -y|0\rangle-x|1\rangle$& $\sigma_z\sigma_x\sigma_z$ \\
           $(|\Phi_3\rangle, |\Phi_2\rangle),$ & $|\Phi_2\rangle_{A_0A_1}$& $|+\rangle_C$ &$ x|0\rangle+y|1\rangle$& $I$ \\
         $(|\Phi_3\rangle, |\Phi_3\rangle)\}$ &$|\Phi_2\rangle_{A_0A_1}$& $|-\rangle_C$ &$ x|0\rangle-y|1\rangle$& $\sigma_z$ \\
          &$|\Phi_3\rangle_{A_0A_1}$& $|+\rangle_C$ &$ x|0\rangle-y|1\rangle$& $\sigma_z$ \\
          &$|\Phi_3\rangle_{A_0A_1}$& $|-\rangle_C$ &$ x|0\rangle+y|1\rangle$& $I$ \\
         \hline
          &$|\Phi_0\rangle_{A_0A_1}$& $|+\rangle_C$ &$ y|0\rangle-x|1\rangle$& $\sigma_x\sigma_z$ \\
           & $|\Phi_0\rangle_{A_0A_1}$& $|-\rangle_C$ &$ y|0\rangle+x|1\rangle$& $\sigma_x$ \\
           $\{(|\Phi_3\rangle, |\Phi_0\rangle),$&$|\Phi_1\rangle_{A_0A_1}$& $|+\rangle_C$ &$ -y|0\rangle-x|1\rangle$& $\sigma_z\sigma_x\sigma_z$ \\
          $(|\Phi_3\rangle, |\Phi_1\rangle),$&$|\Phi_1\rangle_{A_0A_1}$& $|-\rangle_C$ &$ -y|0\rangle+x|1\rangle$& $\sigma_z\sigma_x$ \\
           $(|\Phi_2\rangle, |\Phi_2\rangle),$ & $|\Phi_2\rangle_{A_0A_1}$& $|+\rangle_C$ &$ x|0\rangle-y|1\rangle$& $\sigma_z$ \\
         $(|\Phi_2\rangle, |\Phi_3\rangle)\}$ &$|\Phi_2\rangle_{A_0A_1}$& $|-\rangle_C$ &$ x|0\rangle+y|1\rangle$& $I$ \\
          &$|\Phi_3\rangle_{A_0A_1}$& $|+\rangle_C$ &$ x|0\rangle+y|1\rangle$& $I$ \\
          &$|\Phi_3\rangle_{A_0A_1}$& $|-\rangle_C$ &$ x|0\rangle-y|1\rangle$& $\sigma_z$ \\
         \hline
    \end{tabular}
    \caption{List of Bob's unitary operators corresponding to the other parties measurement outcomes}
    \label{tab:my_label_B}
\end{table}
\begin{table}[ht]
    \centering
    \begin{tabular}{|c|c|c|c|c|}
    \hline
    Mentor's&Bob's & Controller's & State of &Unitary operation \\
   Outcomes& Outcomes&  Outcomes&  Alice's qubit ($A_2$)& by Alice $(U_{Alice})$\\
    \hline
           $\{(|\Phi_0\rangle, |\Phi_0\rangle),$&$|\xi_0\rangle_{B_2}$& $|+\rangle_C$ &$ a|0\rangle+b|1\rangle$& $I$ \\
          $(|\Phi_1\rangle, |\Phi_0\rangle),$&$|\xi_0\rangle_{B_2}$& $|-\rangle_C$ &$ a|0\rangle-b|1\rangle$& $\sigma_z$ \\
           $(|\Phi_2\rangle, |\Phi_1\rangle),$  & $|\xi_1\rangle_{B_2}$& $|+\rangle_C$ &$ b|0\rangle-a|1\rangle$& $\sigma_x\sigma_z$ \\
         $(|\Phi_3\rangle, |\Phi_1\rangle)\}$  &$|\xi_1\rangle_{B_2}$& $|-\rangle_C$ &$ b|0\rangle+a|1\rangle$& $\sigma_x$ \\
         \hline
         $\{(|\Phi_0\rangle, |\Phi_1\rangle),$&$|\xi_0\rangle_{B_2}$& $|+\rangle_C$ &$ a|0\rangle-b|1\rangle$& $\sigma_z$ \\
          $(|\Phi_1\rangle, |\Phi_1\rangle),$&$|\xi_0\rangle_{B_2}$& $|-\rangle_C$ &$ a|0\rangle+b|1\rangle$& $I$ \\
           $(|\Phi_2\rangle, |\Phi_0\rangle),$  & $|\xi_1\rangle_{B_2}$& $|+\rangle_C$ &$ b|0\rangle+a|1\rangle$& $\sigma_x$ \\
          $(|\Phi_3\rangle, |\Phi_0\rangle)\}$ &$|\xi_1\rangle_{B_2}$& $|-\rangle_C$ &$ b|0\rangle-a|1\rangle$& $\sigma_x\sigma_z$ \\
         \hline
         $\{(|\Phi_0\rangle, |\Phi_2\rangle),$&$|\xi_0\rangle_{B_2}$& $|+\rangle_C$ &$ b|0\rangle+a|1\rangle$& $\sigma_x$ \\
          $(|\Phi_1\rangle, |\Phi_2\rangle),$&$|\xi_0\rangle_{B_2}$& $|-\rangle_C$ &$ -b|0\rangle+a|1\rangle$& $\sigma_z\sigma_x$ \\
           $(|\Phi_2\rangle, |\Phi_3\rangle),$  & $|\xi_1\rangle_{B_2}$& $|+\rangle_C$ &$ -a|0\rangle+b|1\rangle$& $\sigma_x\sigma_z\sigma_x$ \\
          $(|\Phi_3\rangle, |\Phi_3\rangle)\}$ &$|\xi_1\rangle_{B_2}$& $|-\rangle_C$ &$ a|0\rangle+b|1\rangle$& $I$ \\
         \hline
          $\{(|\Phi_0\rangle, |\Phi_3\rangle),$&$|\xi_0\rangle_{B_2}$& $|+\rangle_C$ &$ b|0\rangle-a|1\rangle$& $\sigma_x\sigma_z$ \\
          $(|\Phi_1\rangle, |\Phi_3\rangle),$&$|\xi_0\rangle_{B_2}$& $|-\rangle_C$ &$ -b|0\rangle-a|1\rangle$& $\sigma_z\sigma_x\sigma_z$ \\
            $(|\Phi_2\rangle, |\Phi_2\rangle),$ & $|\xi_1\rangle_{B_2}$& $|+\rangle_C$ &$ -a|0\rangle-b|1\rangle$& $\sigma_x\sigma_z\sigma_x\sigma_z$ \\
          $(|\Phi_3\rangle, |\Phi_2\rangle)\}$ &$|\xi_1\rangle_{B_2}$& $|-\rangle_C$ &$ a|0\rangle-b|1\rangle$& $\sigma_z$ \\
         \hline
    \end{tabular}
    \caption{List of Alice's unitary operators corresponding to the other parties measurement outcomes}
    \label{tab:my_label_A}
\end{table}
\section{Preparation of the entangled channels}\label{Preparation of the entangled channels}
Choosing and generating an entangled channel is one of the important parts of a quantum communication protocol. In this section, we will describe the preparation of the quantum channel described in eq(2).
\subsection{Generation of $|\Xi_1\rangle_{M-A}$}
Step 1: Prepare four qubit quantum states with $|0\rangle$ as the initial state of each qubit that is
\begin{equation*}
    |P_0\rangle=|0\rangle_{q_0}\otimes|0\rangle_{q_1}\otimes|0\rangle_{q_2}\otimes|0\rangle_{q_3}
\end{equation*}
Step 2: Apply Hadamard gate (H) on qubits $q_0$ and $q_1$, then the state $ |P_0\rangle$ becomes
\begin{equation*}
\begin{split}
    |P_1\rangle=&\frac{1}{\sqrt{2}}(|0\rangle+|1\rangle)_{q_0}\otimes\frac{1}{\sqrt{2}}(|0\rangle+|1\rangle)_{q_1}\otimes|0\rangle_{q_2}\otimes|0\rangle_{q_3}\\
    =&\frac{1}{2}(|0000\rangle+|0100\rangle+|1000\rangle+|1100\rangle)_{q_0 q_1 q_2 q_3}
\end{split}
\end{equation*}
Step 3: Now CNOT gate is applied on the qubits pairs $(q_0, q_2)$ and $(q_1, q_3)$ with control qubits $q_0$ and $q_1$  and qubits $q_2$ and $q_3$ as target qubit. Then the state $|P_1\rangle$ changes to 
\begin{equation*}
\begin{split}
    |P_2\rangle=&\frac{1}{2}(|0000\rangle+|0101\rangle+|1010\rangle+|1111\rangle)_{q_0 q_1 q_2 q_3}
\end{split}
\end{equation*}
Step: 4 Finally apply control Z-gate (CZ) with qubit $q_0$ as the control qubit and qubit $q_1$ as the target qubit. Then the state   $|P_2\rangle $ reduced to the desired quantum state $|\Xi_1\rangle_{M-A}$.
\begin{equation*}
\begin{split}
    |\Xi_1\rangle_{M-A}=&\frac{1}{2}(|0000\rangle+|0101\rangle+|1010\rangle-|1111\rangle)_{q_0 q_1 q_2 q_3}
\end{split}
\end{equation*}

\subsection{Generation of $|\Xi_2\rangle_{M-B-C}$}
Step 1: Prepare five-qubit quantum states with $|0\rangle$ as the initial state of each qubit that is
\begin{equation*}
    |Q_0\rangle=|0\rangle_{q_0}\otimes|0\rangle_{q_1}\otimes|0\rangle_{q_2}\otimes|0\rangle_{q_3}\otimes|0\rangle_{q_4}
\end{equation*}
Step 2: Apply Hadamard gate (H) on qubits $q_0$ and $q_1$, then the state $ |Q_0\rangle$ becomes
\begin{equation*}
\begin{split}
    |Q_1\rangle&\frac{1}{2}(|00000\rangle+|01000\rangle+|10000\rangle+|11000\rangle)_{q_0 q_1 q_2 q_3q_4}
\end{split}
\end{equation*}
Step 3: Now CNOT gate is applied on the qubits pairs $(q_0, q_2)$, $(q_0, q_4)$, $(q_1, q_3)$ and $(q_1, q_4)$ with control qubits $q_0$ and $q_1$  and qubits $q_2$, $q_2$ and $q_4$ as target qubit. Then the state $|Q_1\rangle$ changes to 
\begin{equation*}
\begin{split}
    |Q_2\rangle=&\frac{1}{2}(|00000\rangle+|01011\rangle+|10101\rangle+|11110\rangle)_{q_0 q_1 q_2 q_3q_4}
\end{split}
\end{equation*}
Step: 4 Finally Apply control Z-gate (CZ) with qubit $q_0$ as the control qubit and qubit $q_1$ as the target qubit. Then the state   $|Q_2\rangle $ reduced to the desired quantum state $|\Xi_2\rangle_{M-B_C}$.
\begin{equation*}
\begin{split}
    |\Xi_1\rangle_{M-B-C}=&\frac{1}{2}(|00000\rangle+|01011\rangle+|10101\rangle-|11110\rangle)_{q_0 q_1 q_2 q_3q_4}
\end{split}
\end{equation*}
\begin{figure}
     \centering
     \begin{subfigure}[b]{0.45\textwidth}
         \centering
         \includegraphics[width=\textwidth]{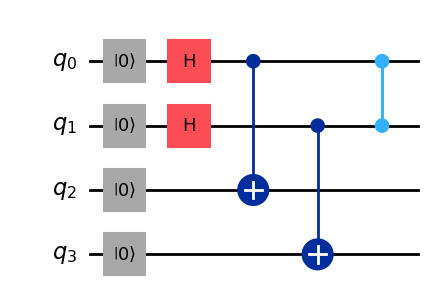}
         \caption{Quantum circuit for $|\Xi_1\rangle_{M-A}$}
         \label{fig:y equals x}
     \end{subfigure}
     \hfill
     \begin{subfigure}[b]{0.45\textwidth}
         \centering
         \includegraphics[width=\textwidth]{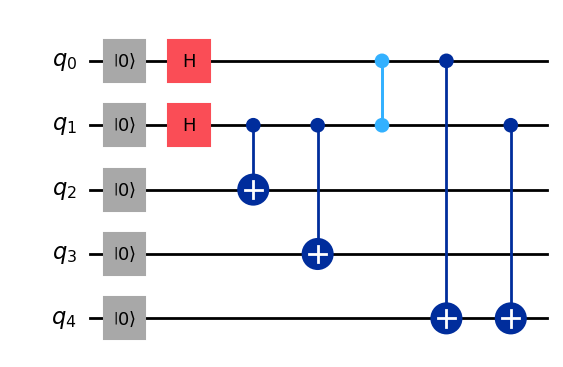}
         \caption{Quantum circuit for of $|\Xi_2\rangle_{M-B-C}$}
         \label{fig:three sin x}
     \end{subfigure}
        \caption{Quantum circuit for the generation of entangled channels}
        \label{fig:three graphs}
\end{figure}
\section{Noise Analysis}\label{Noise Analysis}
The effect of quantum noise is an unavoidable phenomenon in quantum communication protocols. When a qubit is distributed from the source of creation to another party involved in the process quantum noise comes into effect. Suppose The Mentor creates two quantum channels $|\Xi_1\rangle_{M-A}$ and $|\Xi_2\rangle_{M-B-C}$ at his location. The Mentor then distributes the qubits: $A_1$ and $A_2$ are sent to Alice, $B_1$ and $B_2$ to Bob, and $C$ to the Controller. Consequently, the qubits $m_1$, $m_2$, $m_3$, and $m_4$, which remain with the Mentor, are not subjected to environmental noise. However, quantum noise affects the qubits $A_1$, $A_2$, $B_1$, $B_2$, and $C$, during transmission and handling.\\

\noindent The effect of quantum noise on a quantum state is described by the Kraus operator. Suppose a set of Kraus operators $\{K_i: \sum_i K_i=I\}$ describe a noisy environment, then a single qubit quantum state $\rho$ becomes a mixed state which is given by a transformation described as
 \begin{equation}
     \epsilon(\rho)=\sum_{i}K_i \rho K_i^{\dagger}
     \label{n1}
 \end{equation}
 In our protocol Mentor creates two entangled states  $|\Xi_1\rangle_{M-A}$ and $|\Xi_2\rangle_{M-B-C}$. Quantum noise will affect qubits $A_1, A_2, B_1, B_2$ and $C$ of the above two entangled states. In this case, the Kraus operators will be in the form $E_{ijklm}=I\otimes I\otimes I\otimes I\otimes K_i\otimes K_j\otimes K_k\otimes K_l\otimes K_m $. The effect of noise on the combined quantum channel $|\tau\rangle$ given in equation (5) is given by
 
  \begin{equation}
     \epsilon(|\tau\rangle\langle \tau|)=\sum_{ijklm}E_{ijklm} (|\tau\rangle\langle \tau|) E_{ijklm}^{\dagger}
     \label{n2}
 \end{equation}
Here $\sum_{ijklm}E_{ijklm}=I$
Now steps 1 to step 5 are performed which are described in sec 2.  Suppose Mentor's measurement results are $|\Phi_i\rangle_{m_1m_3}$ and $|\Phi_j\rangle_{m_2m_4}$, Alice measurement outcome is $|\Phi_k\rangle_{A_0A_1}$, Bob's measurement result is $|\xi_l\rangle_{B_2}$ and Controller's measurement result is   $|\pm\rangle_{C}$. The final reduced state of the qubits $A_2, B_1$ is given by
\begin{equation}
    L^{out}_{ijklm}=Tr_{m_1m_3m_2m_4A_0A_1B_2C}\big[U(|\psi_0\rangle\langle\psi_0|\otimes\epsilon(|\tau\rangle\langle \tau|))U^{\dagger}\big]
    \label{n3}
\end{equation}
Where $Tr_{m_1m_3m_2m_4A_0A_1B_2C}$ denotes the partial trace on the qubits $m_1, m_3, m_2, m_4, A_0, A_1, B_2$, and $C$. The operator $U$ is given as
\begin{align*}
    U=&\{I_{m_1m_3m_2m_4}\otimes I_{A_0A_1}\otimes (U_{Alice})\otimes (U_{Bob})\otimes I_{B_2}\otimes I_{C} \}\\
\times&\{|\Phi_i\rangle_{m_1m_3}\langle\Phi_i|\otimes|\Phi_j\rangle_{m_2m_4}\langle\Phi_j|\otimes |\Phi_k\rangle_{A_0A_1}\langle\Phi_k|\otimes|\xi_l\rangle_{B_2}\langle\xi_l|\otimes|\pm\rangle_{C}\langle\pm|\}
\end{align*}
Now the fidelity of the output state is given by 
\begin{equation}
    F_{ijklm}= \langle \Psi|L^{out}_{ijklm}|\Psi\rangle
\end{equation}
where $|\Psi\rangle=|\psi_0\rangle\otimes|\psi_1\rangle$ is the desired output state.\\

\noindent Average Fidelity of the protocol is calculated as
\begin{equation}\label{n4}
     F= \frac{1}{256}\sum_{i,j, k=0} ^{3}\sum_{l,m=0}^{1} F_{ijklm}
\end{equation}
\subsection{Bit-flip noisy environment}
\begin{figure}[H]
    \centering
    \begin{subfigure}[b]{0.48\textwidth}
        \centering
        \includegraphics[width=\textwidth]{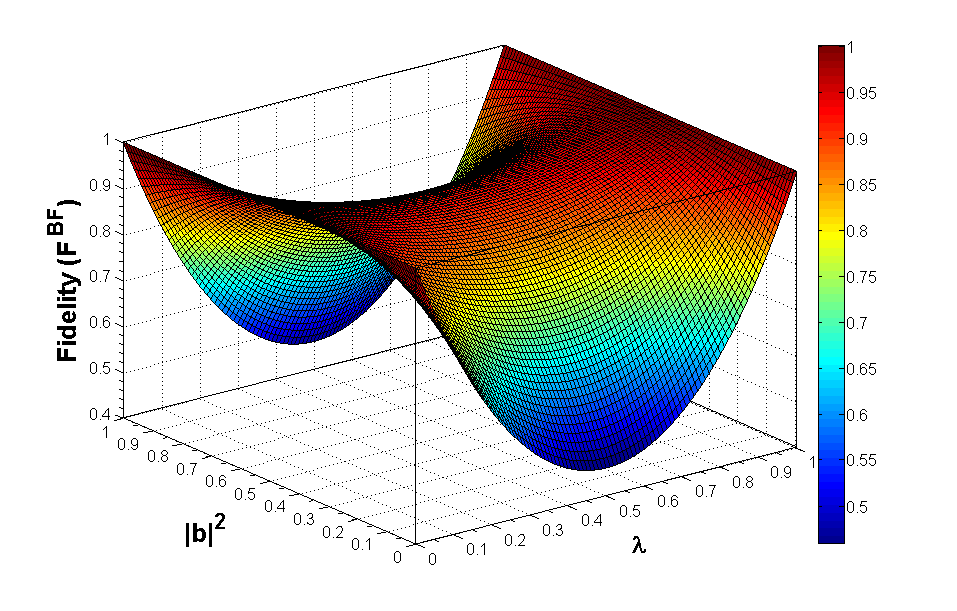}
        \caption{$F$ vs $\lambda$ and $|b|^2$ (Fixed $|y|^2=0.3$)}
    \end{subfigure}
    \hfill
    \begin{subfigure}[b]{0.48\textwidth}
        \centering
        \includegraphics[width=\textwidth]{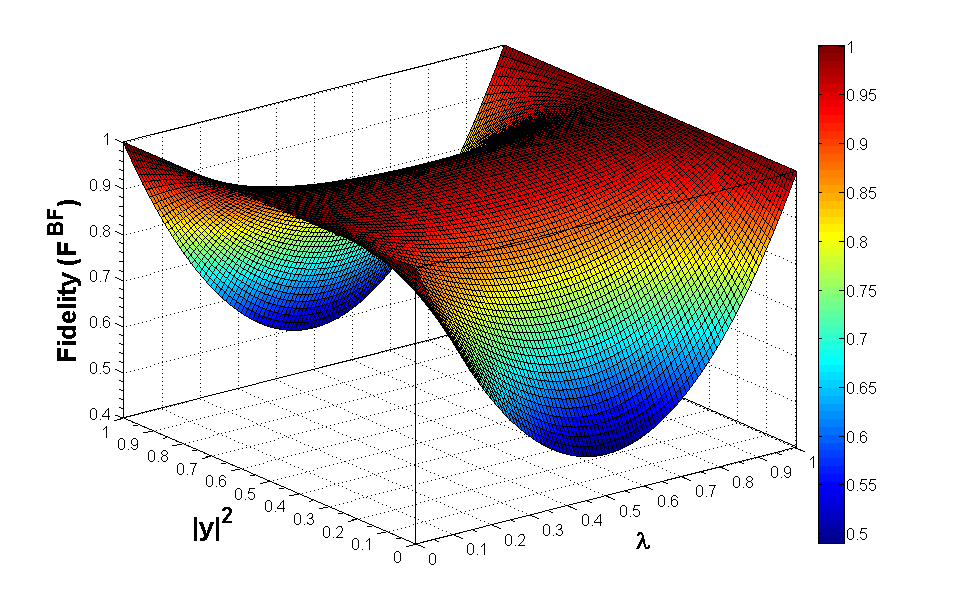}
        \caption{$F$ vs $\lambda$ and $|y|^2$ (Fixed $|b|^2=0.4$)}
    \end{subfigure}
    
    \begin{subfigure}[b]{0.48\textwidth}
        \centering
        \includegraphics[width=\textwidth]{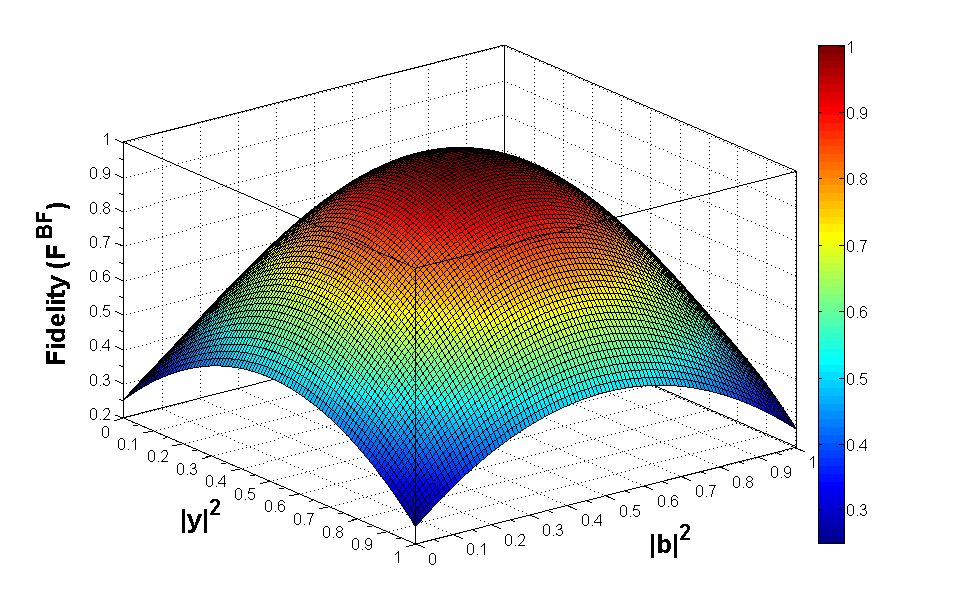}
        \caption{$F$ vs $|b|^2$ and $|y|^2$ (Fixed $\lambda=0.5$)}
    \end{subfigure}
    \hfill
    \begin{subfigure}[b]{0.48\textwidth}
        \centering
        \includegraphics[width=\textwidth]{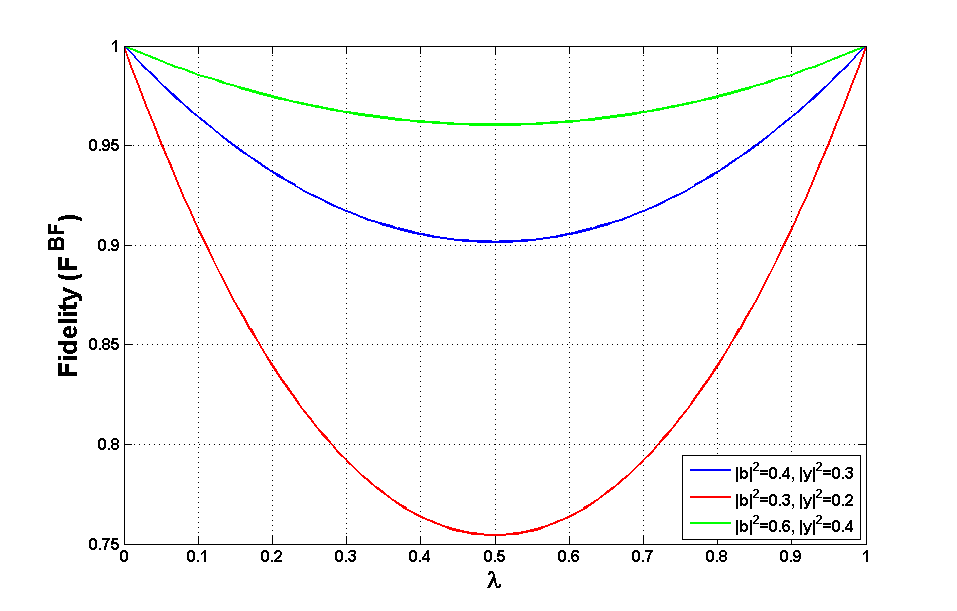}
        \caption{$F$ vs $\lambda$ (Fixed $|b|^2$ and $|y|^2$)}
    \end{subfigure}

    \caption{Plots of Fidelity function $F$ in different parameter variations.}
    \label{fig:fidelity}
\end{figure}

In the presence of bit-flip noise, a qubit remains unchanged with a probability of $(1-\lambda)$ and flips with a probability of $\lambda$. The Kraus operators describing this noise process are given as
\begin{equation}
    \begin{split}
        K_0=
        \begin{pmatrix}
            \sqrt{1-\lambda} & 0\\
            0 & \sqrt{1-\lambda}\\
        \end{pmatrix}
        \text{~and~}
        K_1=
        \begin{pmatrix}
            0 & \sqrt{\lambda}\\
            \sqrt{\lambda} & 0\\
        \end{pmatrix}.
    \end{split}
\end{equation}
Using these Kraus operators, the fidelity of the protocol under the influence of bit-flip noise can be derived through detailed calculations involving equations (\ref{n1}), (\ref{n2}), (\ref{n3}) and (\ref{n4}). The resulting expression for the fidelity is given by
\begin{equation}\label{eq18}
    \begin{split}
        F^{BF}=\{1-2\lambda(1-\lambda)(1-2b^2)^2 \}\{1-2\lambda(1-\lambda)(1-2|y|^2)^2\}.
    \end{split}
\end{equation}
\noindent Variations in the fidelity $F^{BF} $ with respect to different parameters are illustrated in Figure \ref{fig:fidelity}. As observed in Figures \ref{fig:fidelity}(a) and \ref{fig:fidelity}(b), the fidelity behaves similarly when either of the quantum state variables, $|b|^2 $ or $|y|^2 $, is held constant. From Equation (\ref{eq18}) and Figure \ref{fig:fidelity}, we see that the fidelity reaches 1 when the noise parameter is either 0 or 1. This indicates that in the absence of noise, the protocol operates deterministically, and even in the presence of maximum noise, it remains deterministic. The fidelity decreases for $\lambda \in [0, 0.5] $ and increases for $\lambda \in [0.5, 1] $. Moreover, fidelity is equal to 1 when $|b|^2 = |y|^2 \in \{0, 0.5, 1\} $. In particular, when $|\psi_0\rangle_{A_0} = |\psi_1\rangle_{B_0} = \frac{1}{\sqrt{2}} (|0\rangle + |1\rangle) $, the fidelity $F^{BF} $ remains 1, indicating that in this case, it is independent of the noise parameter, meaning that Bit-Flip noise does not affect the protocol.
\subsection{Phase flip noisy environment}
\begin{figure}[H]
    \centering
    \begin{subfigure}[b]{0.48\textwidth}
        \centering
        \includegraphics[width=\textwidth]{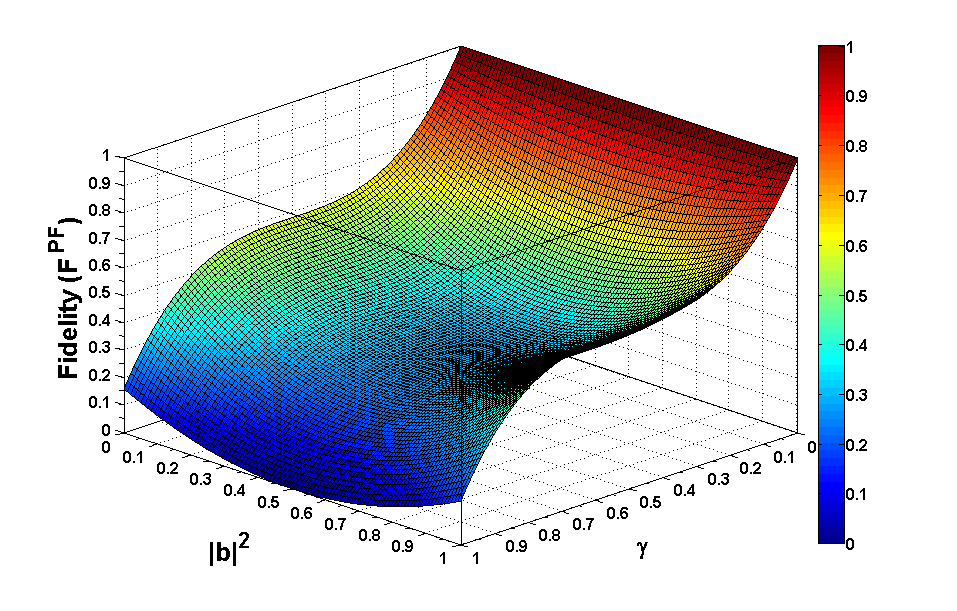}
        \caption{$F^{PF}$ vs $\gamma$ and $|b|^2$ (Fixed $|y|^2=0.3$)}
    \end{subfigure}
    \hfill
    \begin{subfigure}[b]{0.48\textwidth}
        \centering
        \includegraphics[width=\textwidth]{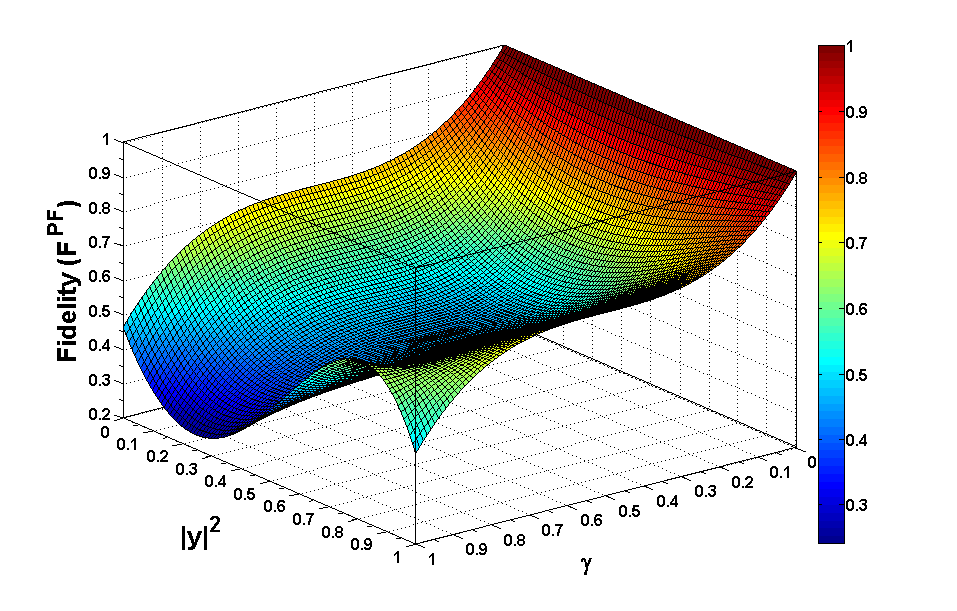}
        \caption{$F^{PF}$ vs $\gamma$ and $|y|^2$ (Fixed $|b|^2=0.4$)}
    \end{subfigure}
    
    \begin{subfigure}[b]{0.48\textwidth}
        \centering
        \includegraphics[width=\textwidth]{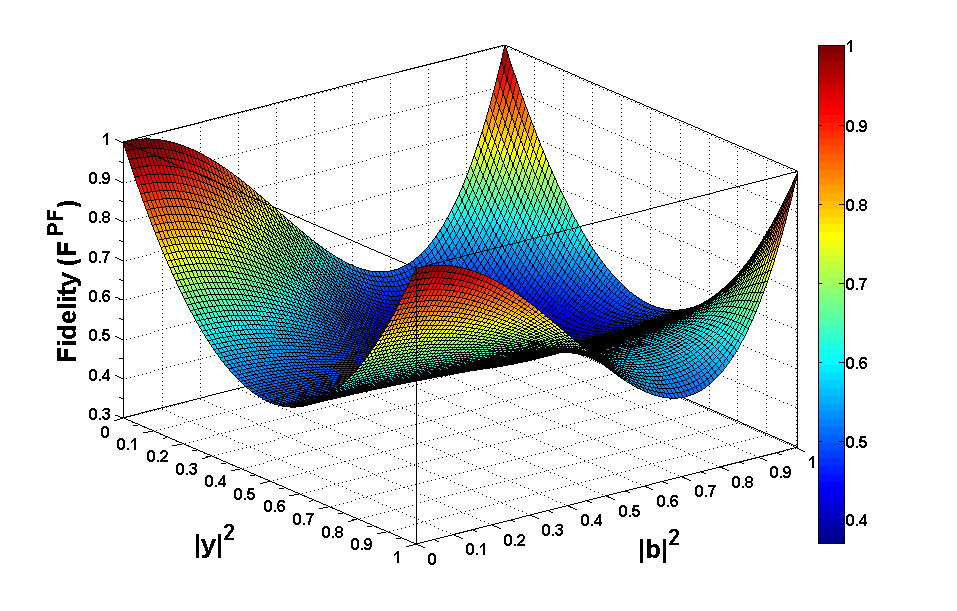}
        \caption{$F^{PF}$ vs $|b|^2$ and $|y|^2$ (Fixed $\gamma=0.5$)}
    \end{subfigure}
    \hfill
    \begin{subfigure}[b]{0.48\textwidth}
        \centering
        \includegraphics[width=\textwidth]{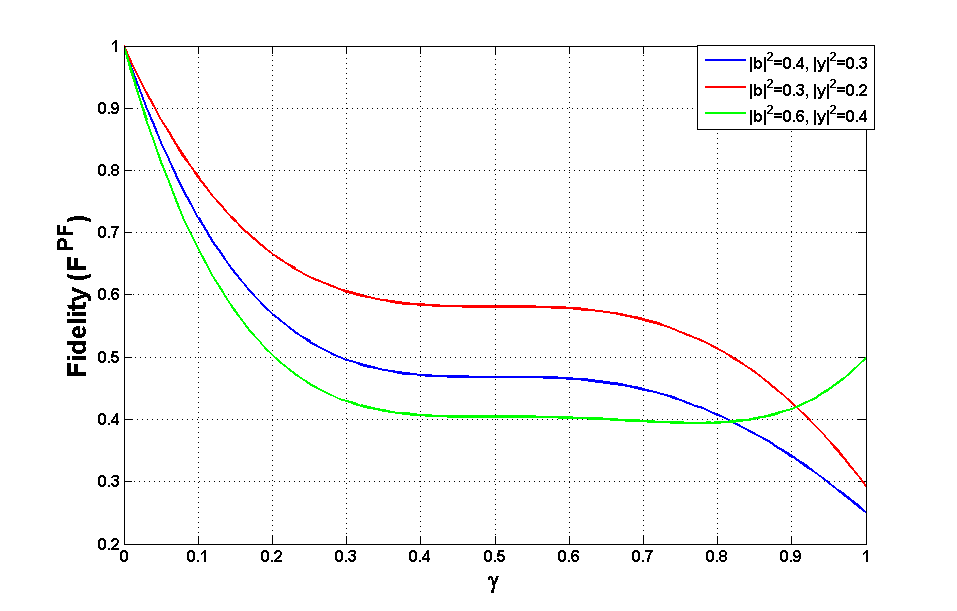}
        \caption{$F^{PF}$ vs $\gamma$ (Fixed $|b|^2$ and $|y|^2$)}
    \end{subfigure}

    \caption{Plots of Fidelity function $F^{PF}$ in different parameter variations.}
    \label{fig:fidelity_PF}
\end{figure}

In a phase-flip noisy environment, the Kraus operators describe how the noise affects a qubit. These operators are given by
\begin{equation}
    \begin{split}
        K_0=
        \begin{pmatrix}
            \sqrt{1-\gamma} & 0\\
            0 & \sqrt{1-\gamma}\\
        \end{pmatrix}
        \text{~and~}
        K_1=
        \begin{pmatrix}
            \sqrt{\gamma} & 0\\
            0 & -\sqrt{\gamma}\\
        \end{pmatrix},
    \end{split}
\end{equation}
where $\gamma$ represents the strength of the phase-flip noise. Using these operators and applying the formulae (\ref{n1}), (\ref{n2}), (\ref{n3}) and (\ref{n4}), the fidelity of the protocol under the effect of phase-flip noise is calculated as
\begin{equation}
    \begin{split}
        F^{PF}=&1-4\gamma|y|^2 (3-6\gamma+4\gamma^2) (1-|y|^2)+\\
        &4(b^4-b^2)\{3\gamma-6\gamma^2+4\gamma^3+4(\gamma-4\gamma^3+4\gamma^4)(|y|^4-|y|^2)\}.\\
    \end{split}
\end{equation}
The variation of fidelity $ F^{PF} $ with respect to the noise parameter $ \gamma $ and the state parameters $ |b|^2 $ and $ |y|^2 $ is shown in Figure \ref{fig:fidelity_PF}. Figure \ref{fig:fidelity_PF}(a) illustrates the dependence of fidelity on $ \gamma $ and $ |b|^2 $ when $ |y|^2 = 0.3 $, demonstrating that fidelity decreases as the noise parameter $ \gamma $ increases. Similarly, Figure \ref{fig:fidelity_PF}(b) depicts the variation of fidelity with $ \gamma $ and $ |y|^2 $ when $ |b|^2 = 0.4 $. When the noise parameter is fixed at $ \gamma = 0.5 $, the variation of fidelity with respect to $ |b|^2 $ and $ |y|^2 $ is shown in Figure \ref{fig:fidelity_PF}(c). Figure \ref{fig:fidelity_PF}(d) presents the fidelity variation with $ \gamma $ for specific initial quantum states. Across all figures, we observe that in the absence of noise ($ \gamma = 0 $), the fidelity remains unity. However, as Phase-Flip noise is introduced, the fidelity of our protocol decreases from unity. 
\subsection{Phase-damping noisy environment}
\begin{figure}[H]
    \centering
    \begin{subfigure}{0.48\textwidth}
        \centering
        \includegraphics[width=\textwidth]{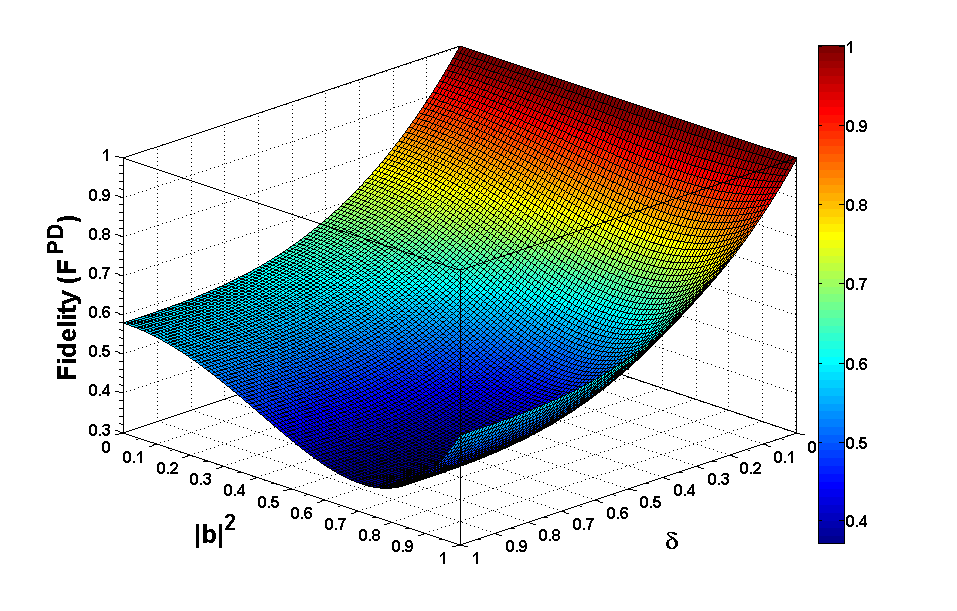}
        \caption{$F^{PD}$ vs $\delta$ and $|b|^2$ (Fix $|y|^2 = 0.3$)}
    \end{subfigure}
    \hfill
     \begin{subfigure}{0.48\textwidth}
        \centering
        \includegraphics[width=\textwidth]{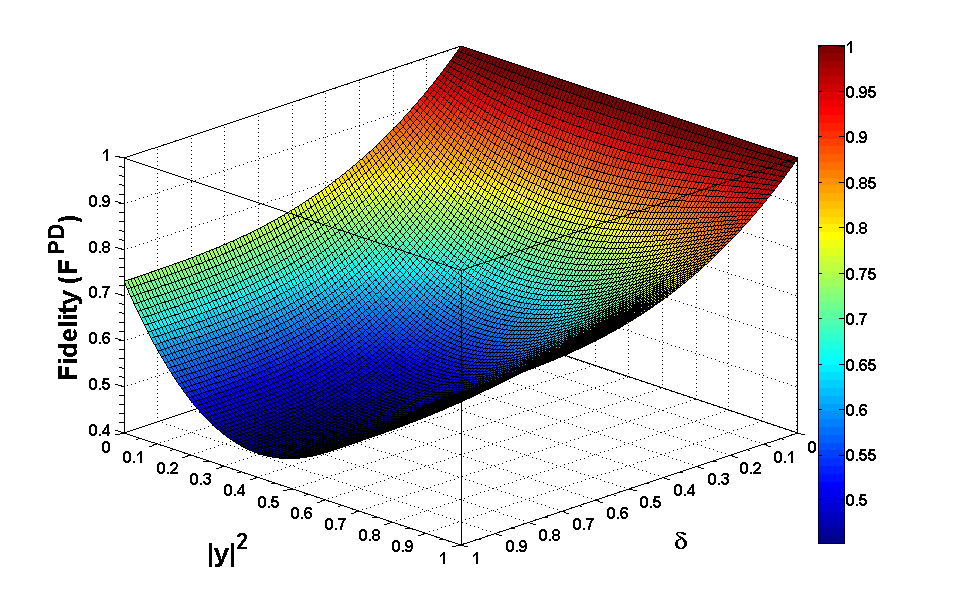}
        \caption{$F^{PD}$ vs $\delta$ and $|y|^2$ (Fix $|b|^2 = 0.4$)}
    \end{subfigure}
    
   
    \begin{subfigure}{0.48\textwidth}
        \centering
        \includegraphics[width=\textwidth]{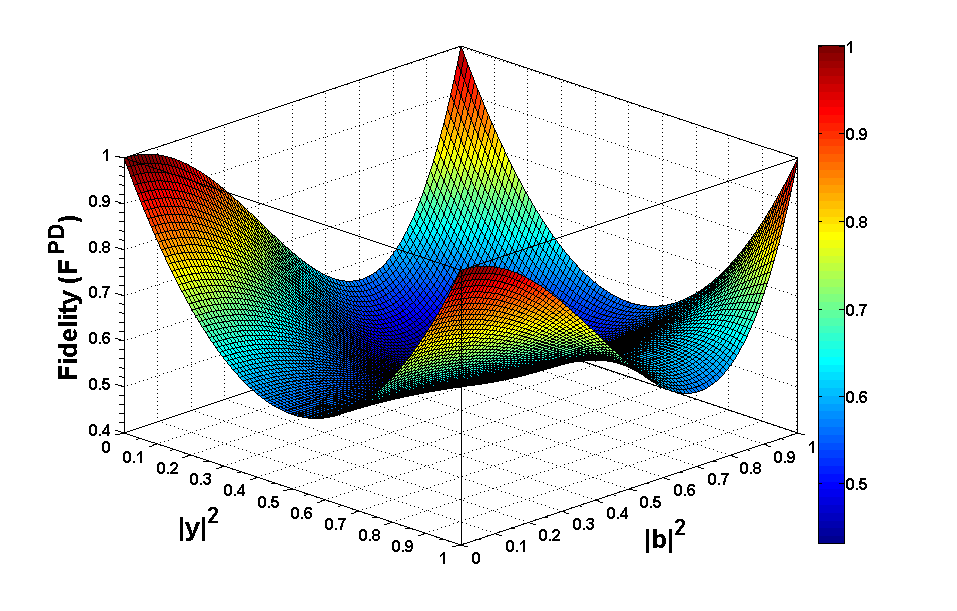}
        \caption{$F^{PD}$ vs $|b|^2$ and $|y|^2$ (Fix $\delta = 0.5$)}
    \end{subfigure}
    \hfill
    \begin{subfigure}{0.48\textwidth}
        \centering
        \includegraphics[width=\textwidth]{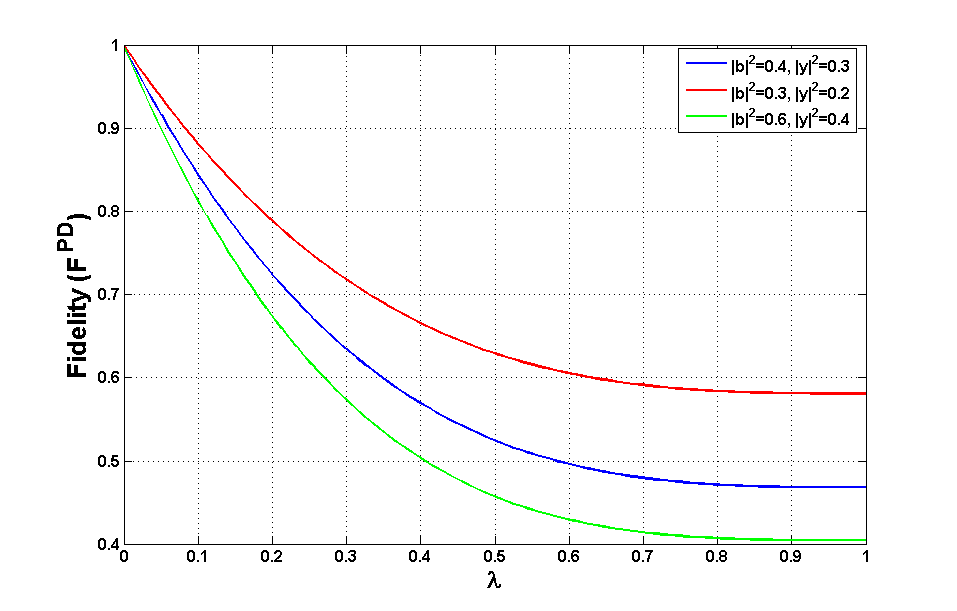}
        \caption{$F^{PD}$ vs $\delta$ (Fixed $|b|^2$ and $|y|^2 $)}
    \end{subfigure}

    \caption{Visualization of $F^{PD}$ under different parameter settings.}
    \label{fig:FPD_plots}
\end{figure}

The Kraus operator for phase-damping noise can be expressed as
\begin{equation}
    \begin{split}
        K_0=
        \begin{pmatrix}
            \sqrt{1-\delta} & 0\\
            0 & \sqrt{1-\delta}\\
        \end{pmatrix},
        K_1=
        \begin{pmatrix}
            \sqrt{\delta} & 0\\
            0 & 0\\
        \end{pmatrix}
        \text{~and~}
        K_2=
        \begin{pmatrix}
            0 & 0\\
            0 & \sqrt{\delta}\\
        \end{pmatrix}
    \end{split}
\end{equation}
where $\delta$ denotes the strength of the phase-damping noise. After performing detailed calculations using equations (\ref{n1}), (\ref{n2}), (\ref{n3}) and (\ref{n4}), the fidelity of the protocol under the influence of phase-damping noise is given by
\begin{equation}
    \begin{split}
        F^{PD}=1+2\delta&\bigg[\{3+(-3+\delta)\delta\}|y|^2(-1+|y|^2)+\\
        &(b^4-b^2)[3-3\delta+\delta^2+2(|y|^4-|y|^2)\{2+(-2+\delta)\delta^2\}]\bigg].\\
    \end{split}
\end{equation}
The variation of fidelity $ F^{PD} $ with respect to the noise parameter $ \delta $ and the state parameters $ |b|^2 $ and $ |y|^2 $ is presented in Figure \ref{fig:FPD_plots}. Figure \ref{fig:FPD_plots}(a) illustrates the dependence of fidelity on $ \delta $ and $ |b|^2 $ while keeping $ |y|^2 = 0.3 $. The figure shows that fidelity decreases as the noise parameter $ \delta $ increases, indicating the detrimental effect of noise on the system.  Similarly, Figure \ref{fig:FPD_plots}(b) depicts the variation of fidelity with $ \delta $ and $ |y|^2 $ when $ |b|^2 = 0.4 $.  When the noise parameter is fixed at $ \delta = 0.5 $, the variation of fidelity with respect to $ |b|^2 $ and $ |y|^2 $ is shown in Figure \ref{fig:FPD_plots}(c). This plot highlights the interplay between the state parameters, revealing regions where fidelity remains relatively high and areas where it rapidly decreases.  Finally, Figure \ref{fig:FPD_plots}(d) presents the fidelity variation with $ \delta $ for specific initial quantum states. Across all figures, it is evident that in the absence of noise ($ \delta = 0 $), the fidelity remains unity, indicating the protocol is deterministic. However, as Phase-Flip noise is introduced, the fidelity decreases from unity, reflecting the degradation of quantum information due to environmental interactions. 
\subsection{Depolarizing noisy environment}
\begin{figure}[H]
    \centering
    \begin{subfigure}[b]{0.48\textwidth}
        \centering
        \includegraphics[width=\textwidth]{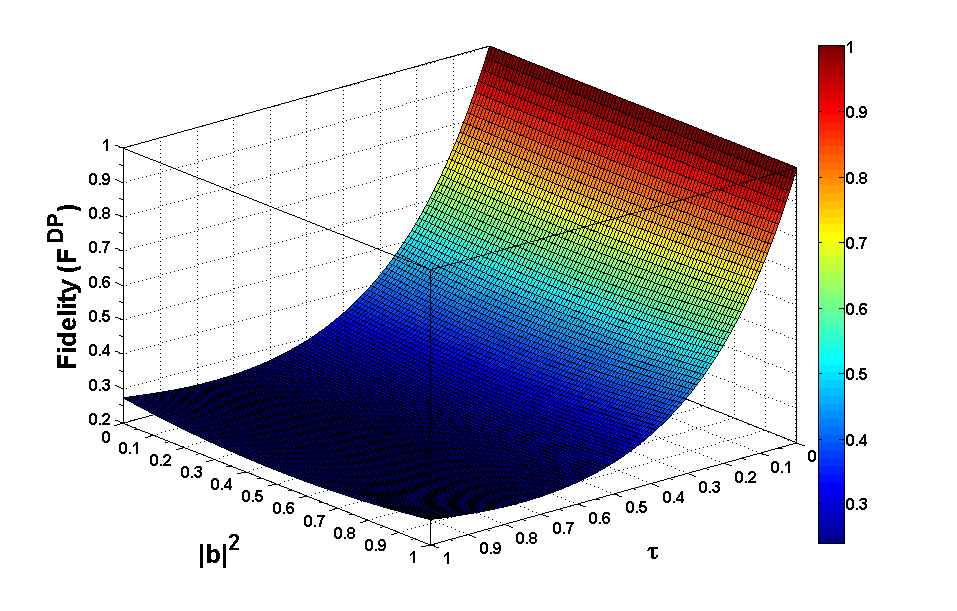}
        \caption{$F^{DP}$ vs. $\tau$ and $|b|^2$ ($|y|^2 = 0.3$)}
        \label{fig:F_tau_b2}
    \end{subfigure} 
     \begin{subfigure}[b]{0.48\textwidth}
        \centering
        \includegraphics[width=\textwidth]{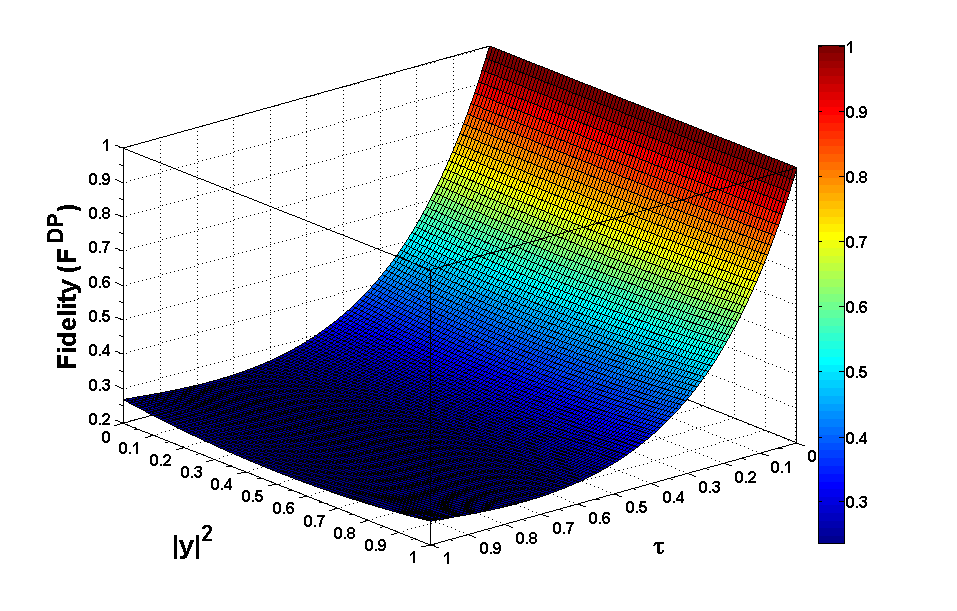}
        \caption{$F^{DP}$ vs. $\tau$ and $|y|^2$ ($|b|^2 = 0.4$)}
        \label{fig:F_tau_y2}
    \end{subfigure}
    
    \begin{subfigure}[b]{0.48\textwidth}
        \centering
        \includegraphics[width=\textwidth]{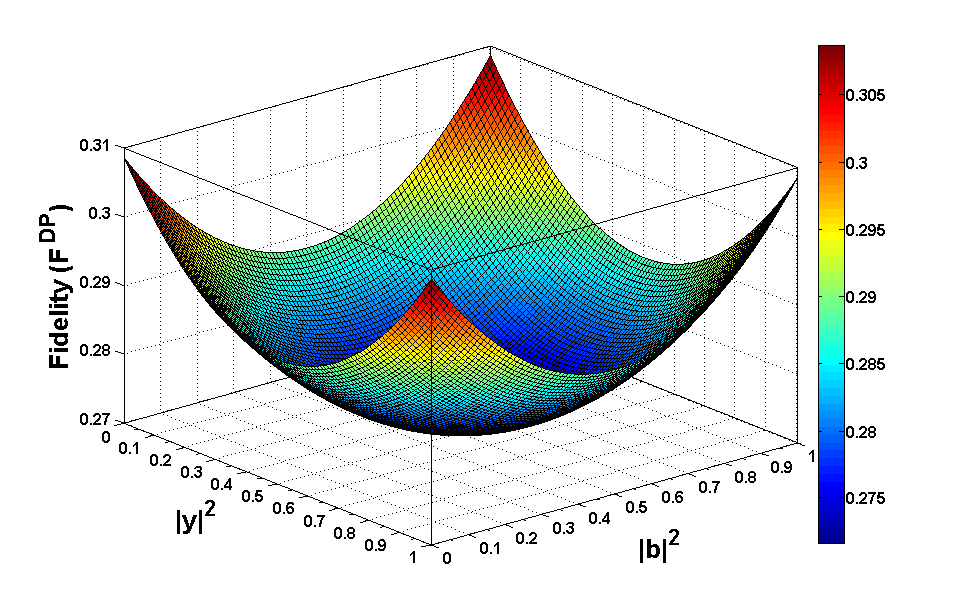}
        \caption{$F^{DP}$ vs. $|b|^2$ and $|y|^2$ ($\tau = 0.5$)}
        \label{fig:F_b2_y2}
    \end{subfigure}
    \begin{subfigure}[b]{0.48\textwidth}
        \centering
        \includegraphics[width=\textwidth]{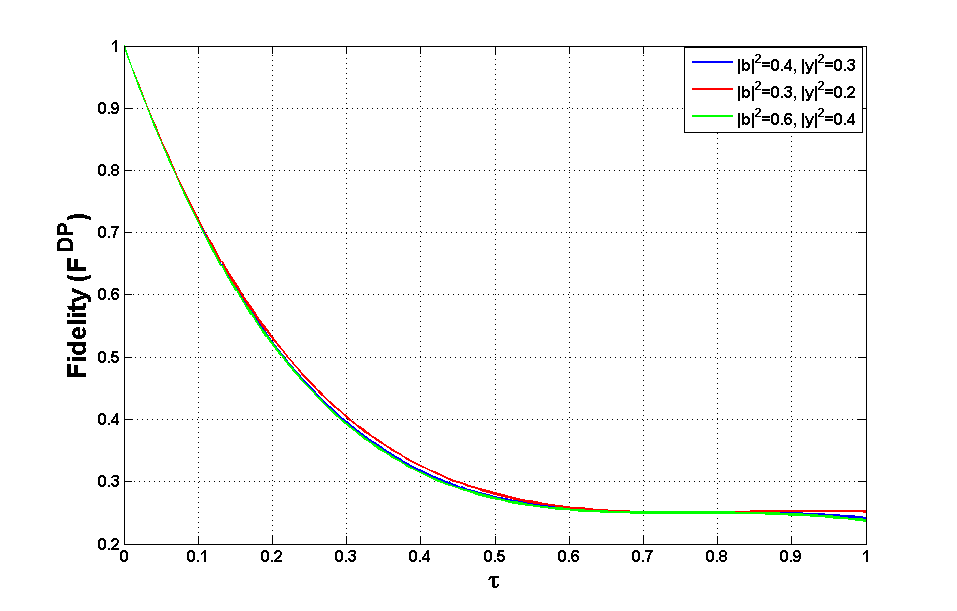}
        \caption{$F^{DP}$ vs. $\tau$ (Fixed $|b|^2$ and $|y|^2 $)}
        \label{fig:F_lambda}
    \end{subfigure}

    \caption{Visualization of $F^{DP}$ for different parameter variations}
    \label{fig:all_FDP}
\end{figure}

The Kraus operator for depolarizing noise can be expressed as
\begin{equation}
    \begin{split}
        K_0=\sqrt{1-\tau}
        \begin{pmatrix}
            1 & 0\\
            0 & 1\\
        \end{pmatrix},
        K_1=\sqrt{\frac{\tau}{3}}
        \begin{pmatrix}
            0 & 1\\
            1 & 0\\
        \end{pmatrix},
        K_2=\sqrt{\frac{\tau}{3}}
        \begin{pmatrix}
            0 & -i\\
            i & 0\\
        \end{pmatrix}
        \text{~and~}
        K_3=\sqrt{\frac{\tau}{3}}
        \begin{pmatrix}
            1 & 0\\
            0 & -1\\
        \end{pmatrix}
    \end{split}
\end{equation}
where $\tau$ is the depolarizing probability.
After carrying out detailed calculations using equations (\ref{n1}), (\ref{n2}), (\ref{n3}) and (\ref{n4}), the fidelity of the protocol in the presence of depolarizing noise is obtained as
\begin{equation}
    \begin{split}
        F^{DP}=1-\frac{8}{243} \tau&\bigg[3(3-2\tau)(9-6\tau+4\tau^2)+(|y|^2-|y|^4)(3-4\tau)^2\{9-4\tau(3-2\tau)\}\\
        &+(b^2-b^4)(3-4\tau)^2\{9-12\tau+8\tau^2-4(3-4\tau)^2 (|y|^2-|y|^4)\}\bigg].
    \end{split}
\end{equation}
The variation of fidelity \( F^{DP} \) with respect to the parameters \( \tau \), \( |b|^2 \), and \( |y|^2 \) is presented in Figure \ref{fig:all_FDP}. These figures provide insight into how the interplay of these parameters affects the fidelity of the system. Figure \ref{fig:F_tau_y2} illustrates the dependence of fidelity on \( \tau \) and \( |y|^2 \), while keeping \( |b|^2 = 0.4 \). The figure shows that fidelity decreases as \( \tau \) increases, indicating a degradation of the system’s coherence with increasing interaction effects. Similarly, Figure \ref{fig:F_tau_b2} depicts the variation of fidelity with \( \tau \) and \( |b|^2 \), with \( |y|^2 \) fixed at 0.3. Figure \ref{fig:F_b2_y2} examines the dependence of fidelity on \( |b|^2 \) and \( |y|^2 \), with \( \tau \) fixed at 0.5. The fidelity remains relatively high when both parameters are small, but as either \( |b|^2 \) or \( |y|^2 \) increases, fidelity decreases significantly. The structure of the function suggests an interplay where certain combinations of \( |b|^2 \) and \( |y|^2 \) can either mitigate or amplify the effects of interaction-induced decoherence. Finally, Figure \ref{fig:F_lambda} presents the fidelity variation as a function of \( \lambda \), with  three different values of \( |b|^2 \) and \( |y|^2  \). The trend shows a monotonic decrease in fidelity as \( \lambda \) increases. Across all figures, a common trend emerges: in the case where $\tau$ tends to zero, fidelity remains close to unity, indicating minimal degradation of the quantum state.
\section{Execution of the protocol in a real IBM quantum computer}\label{IBM}

In this section, we discuss the execution of our protocol on a real quantum computer. Specifically, we have utilized the quantum processing unit (QPU) ibm\_sherbrooke, which was accessible on the IBM Quantum platform \cite{ibm}. As of today, real quantum computers available on the cloud do not support conditional operations, which are unitary operations based on measurement outcomes. Since all the qubits are in one place, we can apply any multiqubit quantum gate. Instead of unitary operations based on different measurement outcomes, we used gates such as CNOT, CZ, and CCNOT. Additionally, in today's real quantum computer systems, the measurement of any qubit is only allowed after all operations are completed.\\

\begin{figure}[ht]
    \centering
    \includegraphics[width=\linewidth]{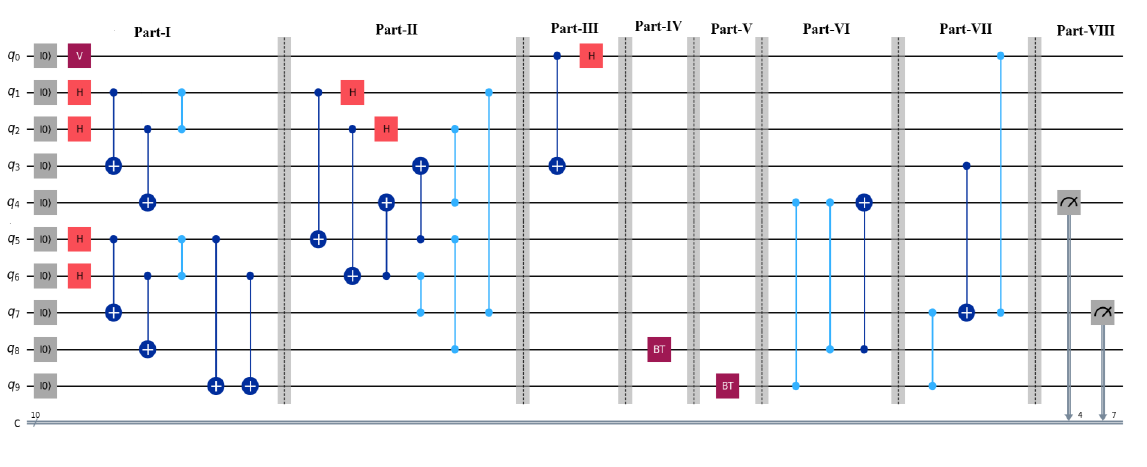}
    \caption{This is the quantum circuit diagram of the proposed protocol.}
    \label{fig:circuit}
\end{figure}

\noindent Suppose Alice wants to teleport an unknown state $|\psi_1\rangle = \frac{1}{\sqrt{3}}|0\rangle + \frac{2}{\sqrt{3}}|1\rangle$  to Bob, while at the same time, Bob wants to create a known state  $|\psi_2\rangle = \frac{1}{\sqrt{2}}|0\rangle + \frac{1}{\sqrt{2}}|1\rangle$ at Alice's location. In this scenario, the quantum circuit for the entire protocol is given in Figure \ref{fig:circuit}, where qubits $q_0, q_1, q_2, q_3, q_4, q_5, q_6, q_7, q_8 $, and $q_9$  represent the qubits  $A_0, m_1, m_2, A_1, A_2, m_3, m_4, B_1, B_2$, and $ C$, respectively. 
The description of the quantum circuit is given as follows:\\

\noindent \textbf{Part I:} This part of the quantum circuit initializes the protocol. Two entangled channels and an unknown state are generated here. To create the quantum state  $|\psi_1\rangle = \frac{1}{\sqrt{3}}|0\rangle + \frac{2}{\sqrt{3}}|1\rangle$ , we use a quantum gate  $V$ , defined as:
$$
V = \frac{1}{\sqrt{3}}
\begin{pmatrix}
    1 & 2 \\
    2 & -1
\end{pmatrix}.
$$
\noindent \textbf{Part II:} This part of the quantum circuit replaces the measurement of the Mentor. All the gates in this section transform the Bell basis measurement into a simple Z-basis measurement.\\

\noindent \textbf{Part III:} This part of the quantum circuit handles the measurement of Alice. Here, we also transform the Bell basis measurement into a simple Z-basis measurement.\\

\noindent \textbf{Part IV:} This part of the quantum circuit handles the measurement of Bob. Since Bob wants to create a known state $|\psi_2\rangle = \frac{1}{\sqrt{2}}|0\rangle + \frac{1}{\sqrt{2}}|1\rangle$  at Alice's location, Bob applies a quantum gate BT, defined as:
$$
BT = \frac{1}{\sqrt{2}}
\begin{pmatrix}
    1 & 1 \\
    1 & -1
\end{pmatrix}.
$$
Here, we also transform the measurement basis  $\{|\xi_0\rangle, |\xi_1\rangle\}$  into a simple Z-basis measurement.\\

\noindent \textbf{Part V:} This part of the quantum circuit handles the measurement of the controller, Candy. Candy applies a quantum gate (BT) to transform the measurement basis  $\{|+\rangle, |-\rangle\}$  into a simple Z-basis measurement.\\

\noindent \textbf{Part VI:} This part of the circuit is for the unitary operations performed by Alice.\\

\noindent \textbf{Part VII:} This part of the circuit is for the unitary operations performed by Bob.\\

\noindent \textbf{Part VIII:} This part of the circuit verifies the protocol by checking if the desired states are teleported/created. In this case, we measure the qubits $ q_4$  and $ q_7$  in the default measurement basis (Z-basis) on the quantum computer.\\

\begin{figure}
     \centering
     \begin{subfigure}[b]{0.45\textwidth}
         \centering
         \includegraphics[width=\textwidth]{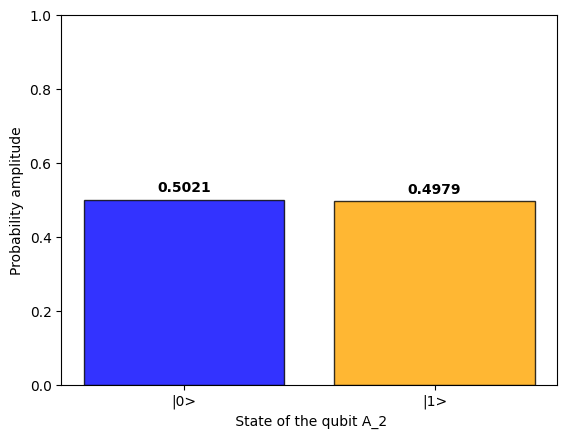}
         \caption{State of the qubit $A_2$}
     \end{subfigure}
     \hfill
     \begin{subfigure}[b]{0.45\textwidth}
         \centering
         \includegraphics[width=\textwidth]{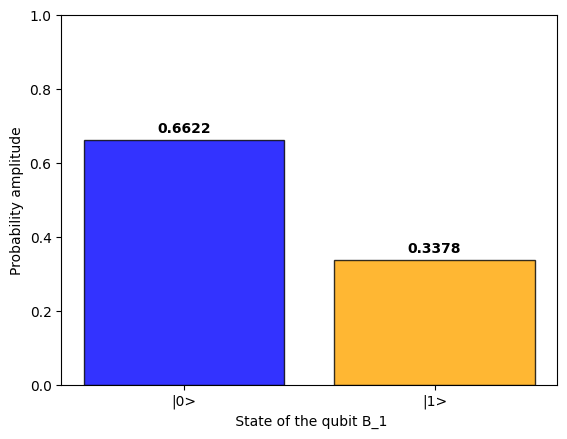}
         \caption{State of the qubit $B_1$}
     \end{subfigure}
        \caption{Output of the qubits $A_2$ and $B_1$}
        \label{output}
\end{figure}

\noindent Finally, we executed the above quantum circuit on the IBM Quantum platform using the QPU ibm\_sherbrooke processor \cite{ibm}, with the number of shots set to 4096. After running the circuit, we recorded the measurement results of qubits $A_2$ and $B_1$, as shown in Fig. \ref{output}. We observe that the outcomes match the desired quantum states.


\section{Generalization of the protocol for n-qubit state}

Suppose Alice possesses an $n$-qubit unknown quantum state in the form $ |\psi_0^n\rangle = (a|00\ldots0\rangle + b|11\ldots1\rangle)_{12\ldots n} $, which she wishes to teleport to Bob. Simultaneously, Bob holds an $m$-qubit known quantum state in the form $ |\psi_1^m\rangle = (x|00\ldots0\rangle + y|11\ldots1\rangle)_{12\ldots m} $, which he intends to create remotely at Alice’s location,  where the coefficients satisfy the normalization conditions $|a|^2+|b|^2=1$ and $|x|^2+|y|^2=1$ The protocol follows the situation outlined in Section 2.\\

\noindent Alice begins by applying CNOT gates to disentangle the qubits in her state. Specifically, she applies CNOT gates to the pairs $ (1,2), (1,3), \ldots, (1,n) $, using qubit$1$ as the control and qubits$2,3,\ldots,n$ as the targets. This operation transforms the state$ |\psi_0^n\rangle $ into:  
$$|\psi_0^n\rangle = (a|0\rangle + b|1\rangle)_1 \otimes |0\rangle_2 \otimes |0\rangle_3 \otimes \cdots \otimes |0\rangle_n. $$

\noindent Alice then teleports qubit $1$ to Bob, while Bob utilizes the protocol described in Section 2 to generate the state $ (x|0\rangle + y|1\rangle) $ at Alice’s location. The entangled channels remain unchanged. After completing the teleportation process outlined in Section 2, Alice and Bob introduce auxiliary qubits in the state$ |0\rangle $.\\

\noindent Alice introduces $m-1$ auxiliary qubits alongside qubit $A_2$, while Bob introduces $n-1$ auxiliary qubits alongside qubit $B_1$. Alice applies CNOT gates to the pairs $ (A_2,2), (A_2,3), \ldots, (A_2,m) $, using $A_2$ as the control qubit and $2,3,\ldots,m$ as targets. This transforms the state into:  
$$|\psi_0^m\rangle = (x|00\ldots0\rangle + y|11\ldots1\rangle)_{A_2 2\ldots m}, $$
which represents Bob’s state.  \\

\noindent Similarly, Bob applies CNOT gates to the pairs $ (B_1,2), (B_1,3), \ldots, (B_1,n) $, using $B_1$ as the control qubit and $ 2,3,\ldots,n$ as targets. This results in:  
$$|\psi_0^n\rangle = (a|00\ldots0\rangle + b|11\ldots1\rangle)_{B_1 2\ldots n}, $$
which completes the teleportation of Alice’s state to Bob. Thus, the task is successfully accomplished.

\section{Discussion and Conclusion}\label{Conclusion}
Hybrid quantum communication refers to the integration of different quantum communication technologies to leverage the strengths of each approach. In this work, we have performed the hybrid quantum communication that combines controlled bidirectional teleportation and remote state preparation which is initiated by a Mentor. The Mentor's task is to create the nine-qubit entanglement channel among the two parties and the controller. The situation is also considered in a realistic scenario where we could not ignore the noise analysis, therefore the effects of various types of noises such as bit-flip, phase-flip, phase-damping, and depolarizing are studied on each qubit of the protocol.\\

\noindent We have calculated the fidelity of the protocol for each of the situations mentioned above. Additionally, we have provided a graphical representation of the variations in fidelity with respect to its parameters. In all cases, when one of the variables $|y|^2$ and $|b|^2$  is fixed, the variation remains largely the same. When the noise parameter is fixed, the variation in fidelity differs in the case of Bit-flip noise compared to other noisy environments. When both initial quantum state parameters are fixed, the variation in fidelity changes with different initial quantum state parameters in the cases of Bit-flip, Phase-flip, and Phase-damping noise. However, in the case of Depolarizing noise, the variation in fidelity remains nearly negligible with a fixed initial state. Moreover, our protocol has been run on a superconductivity-based IBM quantum computer which shows that our communication scheme has viability for real application. In particular, we have utilized the quantum processing unit (QPU) \textbf{ibm\_sherbrooke} \cite{ibm} on the IBM Quantum platform to carry out the protocol implementation.\\

\noindent \textbf{Acknowledgement}  This research work of the first author is supported by the Indian Institute of Engineering Science and Technology, Shibpur and the research work of the third author is supported by the the University Grants Commission (UGC), Government of India (NTA Ref. No.: 211610140697, dated: 19th April 2022).\\


\begin{thebibliography}{99}




\bibitem{teleportation} Bennett, C.H., Brassard, G., Crépeau, C., Jozsa, R., Peres, A., Wootters,W.K., Teleporting an unknown quantum state via dual classical and Einstein–Podolsky–Rosen channels. Phys. Rev. Lett. \textbf{70}, 1895–1899 (1993)
\bibitem{QTM}Sisodia, M., Shukla, A., Thapliyal, K. and Pathak, A., Design and experimental realization of an optimal scheme for teleportation of an n-qubit quantum state. Quantum Information Processing, 16, pp.1-19 (2017).
\bibitem{QT1} Chen, J., Li, D., Liu, M., Yang, Y., Zhou, Q. Quantum controlled teleportation of Bell state using seven-qubit entangled state. Internat. J. Theoret. Phys. 59 (2020).
\bibitem{QT2} Wang, J., Huang, L., Shu, L., Deterministic multi-hop teleportation of arbitrary single-qubit state via partially entangled GHZ-type state. Internat. J. Theoret. Phys. 60(6), 2206–2215 (2021).
\bibitem{QT3} Zhang, Z. and Sang, Y., Bidirectional quantum teleportation in multi-hop communication network. Quantum Information Processing, 22(5), p.201 (2023).
\bibitem{QT4} Ikken, N., Slaoui, A., Ahl Laamara, R. and Drissi, L.B., Bidirectional quantum teleportation of even and odd coherent states through the multipartite Glauber coherent state: Theory and implementation. Quantum Information Processing, 22(10), p.391 (2023).
\bibitem{QT5} Mafi, Y., Kookani, A., Aghababa, H., Barati, M. and Kolahdouz, M.,  Bidirectional quantum controlled teleportation in multi-hop networks: a generalized protocol for the arbitrary n-qubit state through the noisy channel. Quantum Information Processing, 23(10), p.350. 2024.
\bibitem{QT6} Jiang, S.X. and Shi, J., Multi-party three-dimensional asymmetric cyclic controlled quantum teleportation in noisy environment. Quantum Information Processing, 23(7), p.261 (2024).
\bibitem{QT7} Jing, R.H., Huang, Y.B., Yang, J., Bi, A.A., Zhang, J.Y., Xia, K.B. and Zhou, P., Deterministic bidirectional hierarchical teleportation of an arbitrary high-dimensional multi-particle state with a partially entangled quantum channel. The European Physical Journal Plus, 139(10), pp.1-9 (2024).
\bibitem{RSP} Lo, H.K.  Classical-communication cost in distributed quantum-information processing: A generalization of quantum-communication complexity, Phys. Rev. A 62  012313 (2000),
\bibitem{RSP1} Fu, Y., Li, D., Hua, X., Jiang, Y., Zhu, Y., Zhou, J., Yang, X. and Tan, Y., A Scheme for Quantum Teleportation and Remote Quantum State Preparation of Multiple Devices. Sensors, 23(20), p.8475 (2023).
\bibitem{RSP2} Knoll, L.T., Schmiegelow, C.T. and Larotonda, M.A.,  Remote state preparation of a photonic quantum state via quantum teleportation. Applied Physics B, 115(4), pp.541-546 (2014).
\bibitem{RSP3}Kurucz, Z., Adam, P., Kis, Z. and Janszky, J.,  Continuous variable remote state preparation. Physical Review A—Atomic, Molecular, and Optical Physics, 72(5), p.052315 (2005).
\bibitem{RSP4} Ren, S., Han, D., Wang, M. and Su, X., Continuous variable quantum teleportation and remote state preparation between two space-separated local networks. Science China Information Sciences, 67(4), p.142502 (2024). 

\bibitem{JRSP4} Xia, Y., Song, J., and Song, H. S., Multiparty remote state preparation, J. Phys. B: At. Mol. Opt. Phys., 40, 3719 (2007).
\bibitem{JRSP5} An, N. B., and Kim, J., Joint remote state preparation, J. Phys. B At. Mol. Opt. Phys., 41, 095501 (2008).
\bibitem{JRSP1} Choudhury, B. S., and Samanta, S., Perfect joint remote state preparation of arbitrary six-qubit cluster-type states, Quantum Inf. Process., 17, 175 (2018).
\bibitem{JRSP2} Sang, M. H., and Yu, S. D., Controlled joint remote state preparation of an arbitrary equatorial two-qubit state, Internat. J. Theoret. Phys., 58, 2910–2913 (2019).
\bibitem{JRSP3} Peng, J. Y., Bai, M. Q., Tang, L., Yang, Z., and Mo, Z. W., Perfect controlled joint remote state preparation of arbitrary multi-qubit states independent of entanglement degree of the quantum channel, Quantum Inf. Process., 20, 340 (2021).
\bibitem{CRSP1} Wang, Z. Y., Liu, Y. M., Zuo, X. Q., and Zhang, Z. J., Controlled remote state preparation, Commun. Theor. Phys., 52, 235 (2009).
\bibitem{CRSP2} Li, X., and Ghose, S., Analysis of control power in controlled remote state preparation schemes, Internat. J. Theoret. Phys., 56, 667–677 (2017).


\bibitem{CRSP3} Peng, J.Y., Yang, Z., Tang, L. and Bai, M.Q., Controlled remote state preparation of single-particle state under noisy channels with memory, Quantum Information Processing, 22(3), p.145 (2023).
\bibitem{CRSP4} Gu, J.R. and Liu, J.M., Deterministic controlled bidirectional remote state preparation in dissipative environments, Communications in Theoretical Physics, 74(7), p.075101 (2022).
\bibitem{CRSP5} Ma, P.C., Chen, G.B., Li, X.W. and Zhan, Y.B., Cyclic controlled remote state preparation in the three-dimensional system, Laser Physics Letters, 19(11), p.115204 (2022).
\bibitem{CRSP6} Houshmand, M., Jami, S. and Haghparast, M., General controlled cyclic remote state preparations and their analysis, Quantum Information Processing, 23(11), p.372 (2024).
\bibitem{CRSP7} Mandal, M.K., Choudhury, B.S. and Samanta, S., Cyclic controlled remote state preparation protocol initiated by a Mentor for qubits, Optical and Quantum Electronics, 54(9), p.602 (2022).
\bibitem{CRSP8} Shi, J., Controlled cyclic remote state preparation of single-qutrit equatorial states, Modern Physics Letters A, 36(33), p.2150234 (2021).


\bibitem{m1} Choudhury, B.S., Samanta, S. A Remote State Preparation Scheme Initiated and Fixed by a Mentor. Phys. Part. Nuclei Lett. 16, 608–612 (2019).
\bibitem{m2}Mandal, M.K., Choudhury, B.S., Samanta, S. Cyclic controlled remote state preparation protocol initiated by a Mentor for qubits. Opt Quant Electron 54, 602 (2022).
\bibitem{m3} Choudhury, B.S., Mandal, M.K., Samanta, S. Mentor Initiated Controlled Bi-directional Remote State Preparation Scheme For ($2\Leftrightarrow 4$)-Qubit Entangled States in Noisy Channel. Int J Theor Phys 62, 107 (2023).
\bibitem{m4}Mandal, M.K., Choudhury, B.S., Samanta, S. Asymmetric bi-directional teleportation scheme in the presence of a Mentor and a controller. Opt Quant Electron 55, 589 (2023).





\bibitem{H1} Mandal, M.K., Choudhury, B.S. and Samanta, S., Hybrid bidirectional quantum communication protocol of two single-qubit states under noisy channels with memory, Quantum Information Processing, 22(11), p.406 (2023).
\bibitem{H6} Choudhury, B.S., Mandal, M.K., Samanta, S. and Dolai, B., A bidirectional hybrid quantum communication scheme for a known and an unknown qubit, Quantum Studies: Mathematics and Foundations, 10(1), pp.89-99 (2023).
\bibitem{H2} Aharonov, D. K., Stevens, M. L., and Wong, S. Y., Hybrid quantum-classical networks for secure communication, Quantum Information and Computation, 22(9), 1139-1155 (2022).
\bibitem{H3} Smith, J. H., Zhao, L. Z., and Harris, M. P., Quantum repeaters in hybrid communication systems, Nature Communications, 14, 5754 (2023).
\bibitem{H4} Zor, R. K., Davies, N. W., Macek, A. G., et al., Experimental demonstration of hybrid quantum communication over 300 km of fiber, Physical Review Letters, 130(17) (2023).
\bibitem{H5} Cox, H. B., Singh, S. D., and Ford, I. B., Hybrid quantum communication for quantum internet: Challenges and opportunities, npj Quantum Information, 10(1), 17 (2024).
\bibitem{H7} Gong, L., Chen, X.B., Xu, G., Chang, Y., and Yang, Y.X., Multi-party controlled cyclic hybrid quantum communication protocol in noisy environment, Quantum Information Processing, 21(11), p.375 (2022).
\bibitem{H8} Gong, L., Chen, X., Xu, G., and Li, Z., Controlled Cyclic and Bidirectional Hybrid Quantum Communication of Arbitrary Two‐Qubit States, Advanced Quantum Technologies, 7(3), p.2300183 (2024).
\bibitem{H9} Hua, X., Li, D., Fu, Y., Zhu, Y., Jiang, Y., Zhou, J., Yang, X., and Tan, Y., Hierarchical controlled hybrid quantum communication based on six-qubit entangled states in IoT, Sensors, 23(22), p.9111 (2023).
\bibitem{H10} Zhang, J.H. and Jiang, M., Butterfly network coding based on bidirectional hybrid controlled quantum communication, Quantum Information Processing, 21(3), p.107 (2022).





\bibitem{ylzz} Yuan, H., Liu, Y.M., Zhang, W., Zhang, Z.J.: Optimizing resource consumption, operation
complexity and efficiency in quantum-state sharing. J. Phys. B: At. Mol. Opt.
Phys. 41, 145506 (2008).
\bibitem{shy}  Shi, R., Huang, L., Yang, W.: Multi-party quantum state sharing of an arbitrary two-qubit
state with Bell-states. Quant. Inf. Process. 10, 231 - 239 (2011).

\bibitem{ibm} \url{https://quantum.ibm.com/services/resources?resourceType=current-instance&system=ibm_sherbrooke}
\end{thebibliography}
\end{document}